\documentclass[twocolumn,aps,pra,superscriptaddress,amsfonts,longbibliography]{revtex4-1}
\usepackage[colorlinks,linkcolor=blue,urlcolor=red,citecolor=red]{hyperref}
\setcitestyle{super}
\usepackage{graphicx,amsmath,amsfonts,amssymb,capt-of}
\usepackage[tight]{subfigure}
\usepackage{siunitx}
\usepackage{bigints}
\usepackage{tabularx}
\usepackage{mathtools}
\usepackage{float}
\usepackage{multirow}
\usepackage{newfloat}
\usepackage{array}
\usepackage{tikz}
\usetikzlibrary{decorations.pathreplacing}
\usetikzlibrary{colorbrewer}
\usetikzlibrary{arrows}
\usetikzlibrary{arrows.meta}
\usetikzlibrary{calc}
\usetikzlibrary{positioning}
\usepackage{makecell}
\usepackage{comment}
\usepackage{physics}
\usepackage[utf8]{inputenc}
\usepackage{amsmath}

\usepackage{geometry}
\usepackage{xcolor}
\usepackage{bbm}
\newcolumntype{P}[1]{>{\centering\arraybackslash}p{#1}}

\usepackage{enumitem}
\usepackage{chemfig}
\usepackage{circuitikz}
\usepackage{qcircuit}
\usepackage[normalem]{ulem}
\newcolumntype{x}[1]{>{\centering\let\newline\\\arraybackslash\hspace{0pt}}p{#1}}
\DeclareFloatingEnvironment[
    fileext=loa,
    listname=List of Algorithms,
    name=ALGORITHM,
    placement=tbhp,
]{algorithm}

\makeatletter
\newcommand{\vast}{\bBigg@{4}}
\newcommand{\Vast}{\bBigg@{5}}
\newcommand{\VAst}{\bBigg@{6}}
\makeatother

\newenvironment{packed_enum}{
 \begin{enumerate}
   \setlength{\leftmargin}{2em}
   \setlength{\topsep}{1pt}
   \setlength{\partopsep}{1pt}
   \setlength{\itemsep}{1pt}
   \setlength{\parskip}{0pt}
   \setlength{\parsep}{0pt}
 }{\end{enumerate}}

\begin{document}

\normalem
\title{Graph-Theoretic Neural Network Fragmentation with Covariant Direct Molecular Force Learning: Enabling Coupled-Cluster Accuracy AIMD for Fluxional Systems}
\author{Xiao Zhu, Srinivasan S. Iyengar\email{Corresponding Author: iyengar@iu.edu},}
\affiliation{Department of Chemistry, Department of Physics, and the Indiana University Quantum Science and Engineering Center (IU-QSEC), Indiana University, 800 E. Kirkwood Ave, Bloomington, IN-47405}
\date{\today}
\begin{abstract}
Accurate ab initio molecular dynamics (AIMD) simulations of complex, fluxional chemical systems are severely limited by the high computational scaling of correlated electronic structure methods. To overcome this bottleneck, we present a robust, graph-theoretic molecular fragmentation framework integrated with machine learning to directly model post-Hartree-Fock nuclear forces at coupled cluster accuracy. Bypassing the limitations of automatic differentiation on learned energy surfaces that may struggle with link-atom Jacobians, our approach directly predicts nuclear force vectors. By projecting these vectors onto fragment-fixed principal axes of inertia, we establish covariant descriptors that naturally preserve rotational, translational, and permutational invariance. The methodology achieves exceptional high parameter efficiency through a vector-valued training protocol that reduces trainable parameters by over an order of magnitude, while an unsupervised mini-batch $k$-means space tessellation algorithm constructs highly representative training databases using only 10\% to 20\% of reference configurations. We rigorously validated this framework on the highly fluxional solvated Zundel cation ($H_{13}O_{6}^+$). Our fully machine-learning-predicted AIMD trajectories successfully reproduced complex dynamical signatures and key structural characteristics, including radial distribution functions and the velocity autocorrelation power spectrum. Ultimately, this scalable, systematically improvable framework bridges the gap between high-level correlated wavefunction theories and long-timescale reactive sampling, laying the foundation for advanced, LLM-inspired transfer learning in modern chemical dynamics simulations. 
\end{abstract}

\maketitle

\section{Introduction}
\label{sec_intro}

Ab initio molecular dynamics (AIMD)\cite{KarplusBO1,Leforestier-BO,CP,admp1,NA-MQC-Barbatti-2018} plays an important role in the study of reactive systems,  \cite{Parrinello-harmonic,metadynamics-1,ADMP-rareevents,ADMP-rare-isotopes} ensemble averages, \cite{Roy-Gordon-AbsSpectrum,McQuarrie} and vibrational properties beyond the harmonic approximation.\cite{iyengarpetersen,iyengar2005dynamical,iyengarprotonatedwater2,me2o2h+xiaohu,lioomenseyler,Scott-proj,li2016grotthuss,sager2017proton,LiTeigeIyengar,qwaimd-sadafp,iyengardayvoth} Despite its accuracy and generality, AIMD requires calculations of the electronic-structure to be performed at every time step, imposing a severe computational restriction that limits accessible system sizes and simulation times. As a result, it is no surprise that most AIMD based chemical analysis studies performed to date are generally carried out using density functional theory. This limitation becomes particularly pronounced for molecular systems with strong polarization and charge delocalization, for which accurate electronic-structure descriptions are especially challenging\cite{Truhlar-functional-review,DFT-vDW-Michaelides}. Although high-level correlated methods such as CCSD(T) can provide benchmark-quality energetics, their steep algebraic scaling, formally $O(N^7)$ with respect to the number of basis functions $N$, makes on-the-fly AIMD calculations computationally prohibitive for large systems.\cite{H5O2+XCcomp,Schaefer:94,DFT-vDW-Michaelides,Truhlar-functional-review,yang-dft-review}


To accelerate AIMD simulations, machine learning (ML) techniques have been increasingly employed to reduce the cost of electronic structure calculation for nuclear force\cite{ml-force-gdml,ml-force-flare,ml-force-mace,BPNN, deepmd, ani}. In many ML force field frameworks, the total potential energy is expressed as a sum of atomic energy contributions, each parametrized by functions of interatomic distance, angles, or more general local descriptors.  These descriptors are designed to satisfy translational, rotational, and permutational symmetries, thereby encoding local atomic environments in a transferable representation. Although some methods like High Dimensional Neural Network Potential\cite{nnp_review1} include force vector in the loss function to improve the quality and smoothness of the learned PES, force predictions are commonly obtained as the partial derivative of the learned potential energy with respect to nuclear coordinates, often using automatic differentiation. This construction enforces consistency between the learned energy and force models. 


Despite these advances in machine learning techniques, major challenges remain. Primarily, 
the local energy terms used in atom-wise decompositions are usually latent quantities and are typically learned from full-system reference energies and forces. Thus, even when the models describe atom-wise contributions, the training data are often generated from expensive calculations on complete molecular configurations. This difficulty becomes especially pronounced in PES construction and molecular dynamics applications: even when the relevant structural variation is localized to a small number of atoms or reaction coordinates, each reference data point may still require an electronic-structure calculation on the full molecular system.
Due to such reasons, 
the largest system studied
to date using
Coupled-Cluster data is $H_9O_4^+$\cite{waters_nnp} (with absolute error of 0.09 kcal/mol). Furthermore, the resultant neural network model shows limited transferability and the mean absolute error for $H_{13}O_6^+$, using this model is found to be 1.32 kcal/mol\cite{waters_transfer_nnp,BPNN,nnp_review1}. It is precisely this transferability problem that was addressed in Ref. \cite{Xiao-LLM}. In contrast to the above listed studies, the machine learning methods developed here for nuclear forces and in Ref. \cite{Xiao-LLM} for potential energy surfaces, resulted in 0.36 kcal/mol accuracy for the Coupled Cluster level potential energy surface for the complex protonated 21-water cluster (H$_{43}$O$_{21}^+$) system. Indeed, in this paper we obtain nuclear forces for $H_{13}O_6^+$ with 10 micro-Hartree/Bohr accuracy with respect to CCSD calculations and show that AIMD calculations are possible, for medium sized water clusters with coupled cluster accuracy.



To arrive at a solution to the challenges described above, we employ graph-theoretic fragmentation to reduce molecular force modeling from a global high-dimensional learning problem to a collection of asynchronous and local lower-dimensional learning problems that are independent of each other and are potentially transferrable to other problems as shown for energies in Ref. \cite{Xiao-LLM}. Here, graphs are used in a fundamentally different way from graph neural networks (GNNs) \cite{mpnn} or MACE \cite{ml-force-mace}. In GNN-based interatomic potentials, nodes typically represent atoms, and edges define local interaction pathways through which atomic features are updated (as in Bayesian networks\cite{Belief-Kikuchi}) by message passing. The resulting learned atomic representations are then used to predict atom-wise energy contributions.

In contrast, our method represents local atomic groups as nodes and interactions between these groups as edges. For example, nodes may include groups of atoms that interact closely, such as functional groups in organic molecules\cite{frag-PFOA}, amino acids in peptide fragments\cite{CGAIMD}, or water molecules in condensed phase problems\cite{fragPBC,frag-PFOA,frag-ML-Xiao,Xiao-LLM}. Higher-order simplexes, such as triangles formed by three mutually connected nodes and tetrahedrons formed by four mutually connected nodes, are then used to encode higher order (many-body) interactions. The local properties associated with these graph objects, including energies and forces, can then be systematically weighted and assembled to yield full-system potential energies and force vectors.

Our graph-based reconstruction of molecular potential energy also suggests a strong connection with large-language-model-type architectures.\cite{llm-pes, trans-attention} In both cases, a complex high-dimensional object is represented through a collection of lower-dimensional units and the relationships among them. In language models, tokens or words are embedded into an abstract linguistic vector space, and contextual representations are built by combining information from other tokens or words. In the present molecular setting, fragments represented by nodes, edges, triangles, and higher-order graph elements play an analogous role. They encode subsystem information, the fragment energies, and the relationships among molecular subsystems. The full molecular PES is then reconstructed by combining these lower-dimensional subsystem contributions according to the graph topology and many-body interaction structure. Following this idea, lower-dimensional graph-based neural networks have been shown to be transferrable from a 51-dimensional molecular system to a 186-dimensional protonated water cluster, enabling a full-dimensional neural-network PES for the protonated 21-water cluster at CCSD-level accuracy.\cite{llm-pes} In this paper we undertake appropriate generalizations that will make full scale AIMD with machine learning and coupled cluster theory possible for large and complex systems.

This paper is organized as follows. In section \ref{sec_graph}, we  summarize our previous work about applying graph theory to molecular fragmentation and 
in section \ref{sec_force}, we introduce our equivariant molecular geometry 
based descriptor and justify our approach of directly training and predicting force as vector target machine learning tasks. Section \ref{sec_result} begins by detailing our AIMD data set and training model based on unsupervised learning mechanisms such as the k-means based space tessellation algorithm for fragment geometry sampling. Following this, the force prediction accuracy for each individual fragment types is evaluated and subsequently combined to assess full system force accuracy. 
Finally, we present fully (machine learning) predicted machine 
trajectories alongside their associated radial distribution functions and velocity autocorrelation functions to validate the structural and dynamical properties generated by the framework in Section \ref{sec_autocor}. Conclusions are given in Section \ref{sec_autocor}.

\section{graph theoretic fragmentation based machine learning for post-Hartree-Fock gradients}
\label{sec_graph}


We first summarize our graph-theoretic approach to machine learning potential energy surfaces before discussing forces. More details can be found in Refs. \cite{frag-ML-Xiao,frag-PFOA,frag-ONIOM-frag-reformulation,SSI-Review1-QC-ES-QN,Xiao-LLM}.

The graph theoretic fragmentation\cite{fragAIMD,fragAIMD-elbo,fragAIMD-CC,CGAIMD,frag-BSSE-AIMD,frag-AIMD-multitop,fragPBC,fragIJQC-review,frag-PFOA,frag-AIMD-multitop-2,Harry-weighted-graphs,frag-ML-Xiao,frag-QC-Harry,frag-TN-Anup,SSI-Review1-QC-ES-QN,frag-QC-2,frag-ONIOM-frag-reformulation,llm-pes} approach involves defining a graph ${\cal G}$ based on the molecular geometry. 
Groups of atoms from   within a molecular system, based on distance, connectivity and chemical properties, are used to designate nodes within a graph. 
For example, nodes may include individual water molecules\cite{frag-ML-Xiao,Xiao-LLM,fragAIMD,fragAIMD-elbo,fragAIMD-CC,frag-AIMD-multitop,frag-AIMD-multitop-2}, or hydrocarbon fragments as in Ref. \cite{frag-PFOA}, or single amino acid monomers as in Ref. \cite{CGAIMD}. 
In all cases, the set of all nodes is labeled as ${\bf V_0}$. These nodes interact with each other where some of these interactions may arise due to chemical connectivity, and other interactions may be due to non-bonded interactions. In all cases, the defined set of nodes are connected to each other based on some predefined distance cutoff to form the set of edges, ${\bf V_1}$.
Together, the sets of nodes and edges define the graph, ${\cal G}$.
However, our formalism goes beyond the definition of nodes and edges in the graph and here, higher order interactions within the system can be systematically captured and represented as higher rank simplexes\cite{dey1997267,adams2008introduction,berger1984affine} 
defined by the graph, such as triangles in set ${\bf V_2}$ for interactions three different node fragments, and tetrahedrons, in set ${\bf V_3}$, for interactions between a set four node fragments, and so on. 
By defining a maximum rank cutoff ${\cal R}$, we obtain a family of sets of fragments using the simplexes created from the graph, and given by
\begin{align}
\left\{ {\bf V_r}
\left\vert r=0, \cdots,  {\cal R} \right. \right\} \equiv \left\{ {\bf V_0}, {\bf V_1}, {\bf V_2}, \cdots, {\bf V_r}, \cdots, {\bf V_{\cal R}} \right\}.
\label{G-powerset-R}
\end{align}
The components ${\bf V_r}$ can be used to systematically reconstruct the full system properties and is a key idea within our formalism. Thus, the molecular system is treated here as a simplicial complex\cite{dey1997267,armstrong2013basic}. 

Using this idea, the target post-Hartree-Fock electronic structure potential energy, $E^{target}$, at the full system nuclear geometry $\bf{\bar{x}}$ can be expressed as a low-level reference energy $E^{ref}({\bf \bar x})$, such as DFT, augmented by a correction term $\Delta E({\bf {\bar x}})$:
\begin{align}
E^{target}({\bf {\bar x}}) = E^{Ref}({\bf {\bar x}}) + \Delta E({\bf {\bar x}}), 
\label{eq_graph-main}
\end{align}
where the correction term is assembled from the individual fragment contributions obtained from the connectivity of the graph,
\begin{align}
\Delta E({\bf {\bar x}})=
\sum_{r=0}^{\cal R}  { \sum_{\alpha_r  \in {\bf V}_r}{\cal M}_{\alpha_r, r}^{\cal R} \;  \Delta E_{\alpha_r,r}({\bf {\bar x}}_{\alpha_r,r})
},
\label{eq_deltaE}
\end{align}
where the terms ${\bf {\bar x}}_{\alpha_r,r}$ represent the geometry of the portion of the molecular system that belongs within the $\alpha_r$-th simplex-rank-$r$ fragments, and the fragment-specific energy difference is defined as:
\begin{align}
     \Delta E_{\alpha_r, r}({\bf {\bar x}}_{\alpha_r,r}) 
     &=  E_{\alpha_r, r}^{target}({\bf {\bar x}}_{\alpha_r,r}) - E_{\alpha_r, r}^{Ref.}({\bf {\bar x}}_{\alpha_r,r}).
     \label{eq_deltaE-2}
\end{align}
Additionally, in Eq.\ref{eq_deltaE},  
\begin{align}
{\cal M}^{\cal R}_{\alpha_r, r} =  \sum_{m\geq r} {(-1)}^{m+r} p_{\alpha_r}^{r,m}
\label{eq_m}
\end{align}
is an over-counting correction from inclusion-exlusion principle, where $p_{\alpha_r}^{r,m}$ accounts for the number of times the $\alpha_r^{th}$ rank-$r$ fragment is included within all rank-$m$ fragments for ${m\geq r}$.
The computational cost is significantly reduced because the input $\bf {\bar x}$ on the left hand side, which represents the full dimensional molecular coordinates, is reduced to ${\bf {\bar x}}_{\alpha_r,r}$ which only contains the fragment coordinated space. 

The expression in Eq. (\ref{eq_graph-main}) has been widely benchmarked over a series of publications. 
Specifically, we have shown that Eq. (\ref{eq_graph-main}) may be used to construct on-the-fly AIMD trajectories\cite{fragAIMD,fragAIMD-elbo,fragAIMD-CC,CGAIMD,frag-BSSE-AIMD} and potential energy surface calculations\cite{frag-AIMD-multitop,frag-AIMD-multitop-2,Harry-weighted-graphs,frag-TN-Anup} for gas-phase as well as condensed-phase systems\cite{fragPBC,frag-PFOA}.
Both extended Lagrangian as well as Born-Oppenheimer based \textit{ab initio} molecular dynamics simulations can be performed at accuracy comparable to CCSD and MP2 levels of theory with DFT-computational cost.\cite{fragAIMD,fragAIMD-elbo,fragAIMD-CC,CGAIMD}
Hence for the first time, in Refs. \onlinecite{fragAIMD-elbo} and \onlinecite{fragAIMD-CC} we presented Car-Parrinello-style dynamics, but with CCSD accuracy. 
Similarly, we have shown how multiple graphical representations of molecular systems can be used simultaneously to construct accurate potential surfaces in agreement with the MP2 and CCSD levels of theory, again at DFT cost.\cite{frag-AIMD-multitop,frag-AIMD-multitop-2,fragIJQC-review,Harry-weighted-graphs,frag-TN-Anup}. 
In Ref. \onlinecite{frag-BSSE-AIMD}, we have also shown that weak interactions (specifically hydrogen bonds) can be accurately captured and efficient approximations to large-basis AIMD trajectories, such as 6-311++G(2df,2pd), can be constructed through computational effort commensurate with much smaller basis set sizes, sets such as 6-31+G(d).
Furthermore, we have also shown in Ref. \onlinecite{fragPBC,frag-PFOA} how condensed-phase simulations on interfaces and liquids may be constructed with hybrid DFT accuracy at gradient-corrected DFT accuracy. We also provide novel approaches to construct reduced circuit depth quantum computing algorithms in Ref. \onlinecite{frag-QC-Harry,frag-QC-2}. Finally, the approach has been demonstrated for hydrogen-bonded systems (such as water clusters and condensed phase systems) and also for covalently bonded biological systems and hydrocarbon where link-atoms\cite{CGAIMD,frag-BSSE-AIMD,frag-PFOA} are included within Eq. (\ref{eq_graph-main}). 
This work is reviewed in Refs. \onlinecite{fragIJQC-review,SSI-Review1-QC-ES-QN,frag-ONIOM-frag-reformulation}.

The expression in Eq. (\ref{eq_graph-main}) is closely related to the ONIOM formalism\cite{oniom,frag-ONIOM-frag-reformulation} where the energy at some lower level is augmented through a correction term. However, through Eq. (\ref{eq_deltaE}), the approach here is also closely related to many-body-expansions\cite{CGAIMD}, but here the many-body contributions are directly computed using the connectivity of the graph. In this sense, the definition of high-order simplexes in Eq. (\ref{G-powerset-R}) turns out to be a key step, since simplexes are defined as rank-$r$ objects within which all pairs of nodes are connected\cite{dey1997267,adams2008introduction,berger1984affine}. This definition helps us to recover the many-body expansion from Eq. (\ref{eq_deltaE}). 

\subsection{Graph theory based machine learning protocols for potential energy surfaces with post-Hartree-Fock accuracy}
\label{sec_graph-A}
As system sizes grow, simplexes sizes also may also grow and then it becomes important to derive a general machine learning protocol to study such systems, and particularly those fragments that are larger in size. This is precisely what was done in Refs. \cite{frag-ML-Xiao,frag-PFOA,llm-pes}. 
To take advantage of the locality of fragment coordinate spaces, we used a distributed set of independent neural networks to approximate fragment energy corrections to $\Delta E_{\alpha_r,r}({\bf {\bar x}}_{\alpha_r,r})$ and combine the resultant predictions for every fragment to yield a full system {\em target} energy prediction.
In this manner, Eq. (\ref{eq_deltaE}) becomes similar to the so-called $\Delta-ML$ protocol\cite{bowman-2021-deltaML,delta_ml_intr,delta_ml_2,delta_ml_3,deltaml_2024}, leading to 
\begin{align}
\Delta E^{ML}({\bf {\bar x}})=
\sum_{r=0}^{\cal R}  { \sum_{\alpha_r  \in {\bf V}_r}{\cal M}_{\alpha_r, r}^{\cal R} \; \Delta E_{\alpha_r,r}^{ML}({\bf {\bar x}_{\alpha_r,r}}).
} 
\label{eq_graph-ML}
\end{align}
and 
\begin{align}
E^{target}_{ML}({\bf {\bar x}}) = E^{Ref}({\bf {\bar x}}) + \Delta E^{ML}({\bf {\bar x}}), 
\label{eq_graph-main-ML}
\end{align}
As noted earlier, 
since $\bf {\bar x}_{\alpha_r,r}$ depends only on the localized fragment coordinate space, we only need to provide fragment level energies and forces as part of the neural network training process, which leads to large reduction in training costs. 

In Ref. \cite{Xiao-LLM}, we have used Eq. (\ref{eq_graph-ML}) to provide a transfer learning strategy thought and isomorphism to the so-called ``attention'' mechanism\cite{trans-attention,softmax_att} in large language models and this aspect is summarized in Appendix \ref{LLM}. Additionally, in Ref. \cite{frag-PFOA} we have employed a hybrid model where the smaller fragments are computed and the larger fragments are extrapolated.

\subsection{Machine learning post-Hartree-Fock energy gradients}
Given the energy expressions in 
Section \ref{sec_graph-A}, we may construct the graph-theoretic approximation 
for 
nuclear forces 
as the gradients derived from the partial derivative of each energy term in Eq. (\ref{eq_graph-main}):
\begin{align}
\frac{\partial E^{target}({\bf {\bar x}})} {\partial{\bf {\bar x}}} \approx \frac{\partial E^{Ref}({\bf {\bar x}})}{\partial{\bf {\bar x}}} + \frac{\partial \Delta E({\bf {\bar x}})}{\partial ({\bf {\bar x}})}
\end{align}
where the exact gradient correction,
\begin{align}
    \Delta F^E({\bf{\bar{x}}}) = \frac{\partial E^{target}({\bf {\bar x}})} {\partial{\bf {\bar x}}} - \frac{\partial E^{Ref}({\bf {\bar x}})}{\partial{\bf {\bar x}}}
\end{align}
may be expressed using gradients of Eq. (\ref{eq_deltaE}) as
\begin{align}
\frac{\partial \Delta E({\bf {\bar x}})}{\partial ({\bf {\bar x}})} 
& = 
\sum_{r=0}^{\cal R}  { \sum_{\alpha_r  \in {\bf V}_r}{\cal M}_{\alpha_r, r}^{\cal R} \;  \frac{\partial \Delta E_{\alpha_r,r}({\bf {\bar x}}_{\alpha_r,r})}{\partial ({\bf {\bar x}}_{\alpha_r,r})}\frac{\partial ({\bf {\bar x}}_{\alpha_r,r})}{\partial ({\bf {\bar x}})}
} .
\label{eq_graph-f}
\end{align}
Note that when chemical bonds need to be cleaved to create fragments\cite{CGAIMD,frag-PFOA}, ${\bf {\bar x}}_{\alpha_r,r}$ may include link atoms\cite{oniom,ONIOM-admp,CGAIMD,frag-ONIOM-frag-reformulation}.  
In those cases, the link atom positions depend on the molecular connectivity and is always a linear transform of atomic positions in $\bf \bar{x}$. Specifically, the link atom is some fraction of the distance between the substituted atom and the connected atom according to \cite{fragAIMD-elbo,ONIOM-admp}
\begin{align}
    \bar r_{link} = \bar r_{bond} + g(\bar r_{sub}–\bar r_{bond})
    \label{eq_link}
\end{align}
where $\bar r_{bond}$ represents the position of the bonded atom in the fragment and $\bar r_{sub}$ represents the position of the substituted connecting atom in the environment. The factor $g$ in Eq. (\ref{eq_link}) is a scalar quantity normally set as the ratio of a typical bond length between the bonded atom and the link $R_{bond, link}$, and a typical bond length between the bonded atom and the substituted atom $R_{bond, sub}$
\begin{align}
    g = \frac{R_{bond, link}}{R_{bond,sub}}.
    \label{eq_link2}
\end{align}
In general, the term $\frac{\partial ({\bf {\bar x}}_{\alpha_r,r})}{\partial ({\bf {\bar x}})}$ in Eq. (\ref{eq_graph-f}) serves as the necessary Jacobian to transform the fragment gradient and arrive at the full system force\cite{fragAIMD,ONIOM-admp,CGAIMD}.
These methods have been successfully employed to construct on-the-fly AIMD trajectories\cite{fragAIMD,fragAIMD-elbo,fragAIMD-CC,CGAIMD,frag-BSSE-AIMD} and potential energy surface \cite{frag-AIMD-multitop,frag-AIMD-multitop-2,Harry-weighted-graphs,frag-TN-Anup} for gas-phase as well as condensed-phase systems\cite{fragPBC,frag-PFOA}. 

As in the case of energy, for force prediction, we have 
\begin{align}
\frac{\partial \Delta E^{ML}({\bf {\bar x}})}{\partial ({\bf {\bar x}})} &\approx \Delta F^{ML}({\bf {\bar x}})\nonumber \\
&=\sum_{r=0}^{\cal R}  { \sum_{\alpha_r  \in {\bf V}_r}{\cal M}_{\alpha_r, r}^{\cal R} \; \Delta F_{\alpha_r,r}^{ML}({\bf {\bar x}_{\alpha_r,r}}) \frac{\partial ({\bf {\bar x}}_{\alpha_r,r})}{\partial ({\bf {\bar x}})}.
} 
\label{eq_graph-ML-f}
\end{align}
However, when link atoms are involved, the Jacobian term,  $\frac{\partial ({\bf {\bar x}}_{\alpha_r,r})}{\partial ({\bf {\bar x}})}$ couples multiple fragments. For example, when two nodes are covalently connected, the bond bridging the two nodes may be cleaved and saturated with link atoms. See Eqs. (\ref{eq_link}) and (\ref{eq_link2}). In such cases, the set ${\bf {\bar x}}_{\alpha_r,r}$ includes additional atoms that are not part of the set represented by ${\bf {\bar x}}$. The gradients due to link atoms are then to be absorbed into the neighboring fragments, including neighboring nodes, and superseding edges, faces, and higher rank simplexes and this is the role of the Jacobian above.

Computing gradients in such cases is confounded by the problem that the neighboring nodes, that were trained for energies through machine learning are now appended by additional contributions due to network connectivity and link atoms on neighboring fragments. This potentially takes these nodes away from their training subspace,  presenting the need for additional training. We addressed a similar problem in Ref. \cite{Xiao-LLM}, however, in this case, fragmentation  methods with broken bonds present additional challenges for pre-trained energy models within automatic differentiation. 
Additionally, while energies are local in ${\bf {\bar x}}$, gradients inherently require some information from the neighborhood of ${\bf {\bar x}}$ and hence obtaining accurate gradients are inherently more challenging. 

In this work we provide a balanced approach for training energies and forces, and for their use in computing AIMD trajectories and for obtaining molecular potential surface. 
While automatic differentiation has been employed to obtain gradients in previous studies\cite{ml-force-mace}, 
in this paper, we define neural networks that directly provide gradient that are then corrected using the Jacobian term as indicated by Eq. (\ref{eq_graph-ML-f}). Associated algorithms are discussed in the next section.

\section{Force vector training as functions of fragment interatomic distances}
\label{sec_force}
As noted above, in many machine learning approaches to molecular potential energy surfaces\cite{ml-force-gdml,ml-force-flare,ml-force-mace,BPNN, deepmd, ani}, nuclear forces are obtained by differentiating a learned neural network energy model with respect to nuclear coordinates. Although the resulting force at a given configuration is evaluated locally, its accuracy depends on whether the learned energy model correctly captures the local curvature of the potential energy surface in the neighborhood of the configuration. 
Therefore, simultaneous prediction of accurate energies and forces places stronger constraints on the learned representation than fitting energies alone, since the model must reproduce not only the scalar energy values but also their variations with respect to nuclear displacements.
Consequently, more training data is generally needed if the goal is to have both energy and forces to be estimated accurately from the same set of neural networks. 

This challenge is amplified in molecular dynamics applications. Because the predicted forces determine the subsequent trajectory, small force errors can accumulate over time and drive the system toward the boundary of the training distribution or into sparsely sampled regions of the potential energy surface. In practice, these configurations often require additional reference calculations for model refinement. When forces are derived from energy models, accuracy of local derivatives in sparsely sampled regions are difficult to maintain. This adds additional constraint to the learning problem and transfer learning process beyond fitting energies and forces at isolated configurations.

Here, we adopt a different strategy. We choose to utilize different neural networks for training energies, $\left\{ \Delta E_{\alpha_r,r}^{ML}({\bf {\bar x}_{\alpha_r,r}}) \right\}$ (Eq. (\ref{eq_graph-ML})) and forces, $\left\{ \Delta F_{\alpha_r,r}^{ML}({\bf {\bar x}_{\alpha_r,r}}) \right\}$ (Eq. (\ref{eq_graph-ML-f})). Force predictions are generated directly from the dictated force models rather than the partial derivative of the predicted energies. The approach needs far less data to train in high dimensional configuration space. In fact, as shown in this section and the following result section, accurate energy, nuclear force predictions and ML-based AIMD generated structural distribution functions and velocity correlation functions can be achieved using only 10\%-20\% of the available training data with an appropriate model architectural design.

In this section, we start by discussing a interactomic-distance-based equivariant descriptor for direct force vector training. Following this, we analyze the mathematical distinction between training models for scalar target quantities, such as the potential energies, and vector-value target quantities, such as nuclear force vectors. We make the case that vector-value training strategies are beneficial here. Finally, a summary of the mathematical formulation is given at the end of this section.


\subsection{Symmetry preserving interatomic distance descriptors for fragment neural networks}\label{sec_des}
A consistent geometrical representation that respects the physical constraints and symmetry properties of molecular fragments is essential for accurately characterizing their local chemical environments and the associated molecular properties. 
For example, the same molecular fragment may be present in different orientations when the data is obtained from multiple snapshots (or even a single snapshot) in a dynamical system. A universal approach to orient these fragments is needed to construct a consistent learning framework. 
Here we introduce a systematic process to generate an interatomic distance based descriptor, as in Ref. \onlinecite{llm-pes}, where we preserve translational, rotational, and permutational invariance. This descriptor is used for training and prediction of both the 
fragment energies and force vectors. Prediction of vector-valued quantities, such as nuclear forces, 
requires fragment orientational information. This information is generated along with the invariant descriptors during the 
geometry preprocessing step,
and also used to define rotation-covariant post-processing operators. Thus, the key steps involved are (a) rotating a fragment geometry to an internal reference frame defined by its mass moment of inertia axis, (b) deriving neural network predictions, and (c) rotating the predictions back to the original coordinate system.
Since operations within (a) and (c) do not parametrize the learning model itself, we exclude these from the definition of descriptors and discuss their construction and application separately below.

\begin{figure}
    \subfigure[Largest moment of inertia axis (x axis)]{\includegraphics[width=0.95\linewidth]{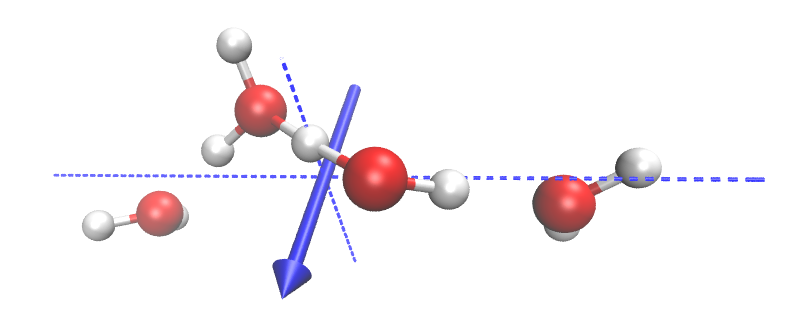}}
    \subfigure[Second largest moment of inertia axis (y axis)]{\includegraphics[width=0.95\linewidth]{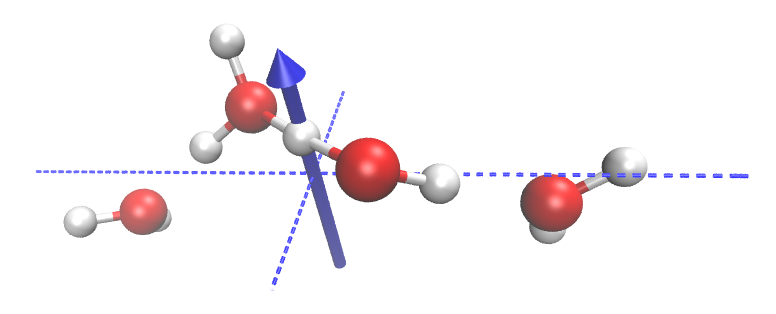}}
    \subfigure[Smallest moment of inertia axis (z axis)]{\includegraphics[width=0.95\linewidth]{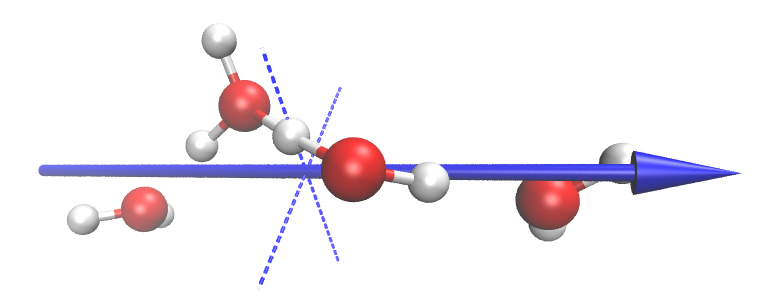}}
    \caption{The three principle mass moment of inertia axis, namely $x, y, z$ for a fragment.}
    \label{fig_principle_axis}
\end{figure}

For a given molecular system or fragment geometry, we record the rotation information and create the descriptor by the following steps:
\begin{packed_enum}
\item We begin by computing the principal axes of inertia for each fragment. An illustration of this step for a fragment containing four water molecules is provided in Figure \ref{fig_principle_axis}.

\item To obtain a consistent atomic ordering for the input coordinate vector $\bf \bar{x}$, atoms are first sorted by atomic number. Atoms of the same element are then ordered according to their respective projections along a chosen principal axis of inertia. For potential energy, we use the axis with the smallest mass moment of inertia to maximize spatial arrangement in Ref.\onlinecite{Xiao-LLM}, but for force, we use all three axes and generates three distinct atomic orderings for the same fragment.

\item For each atomic ordering, an interatomic distance matrix is computed and reshaped into a vector suitable as input to the force neural networks. This input vector is referred to the symbol $\bf{\bar{x}}$ in previous sections. These three input vectors contain the same set of interatomic distances but differ in their ordering due to the choice of principal axis.

\item The three input vectors are used to construct independent learning protocols for the three vector components of the force vector, one along each principal axis. Accordingly, three separate (vector) neural network models are trained, each taking one of the distance vectors as input and predicting the corresponding force vector component for all atoms in the fragment. 
This design of 3 length $N$ vector target prediction will be justify in detail in  Section \ref{sec_vector_learn}.

\item Training data for the three neural networks are generated by projecting the original force vector, $\Delta F_{\alpha_r,r}$, defined in the reference frame of the full system, onto the three principal axes. Each projected force vector component serves as the target output for the corresponding neural network, allowing the force components to be learned separately.


\item During inference, the predicted force components are transformed back into the original reference frame for the fragment by reversing the sequence of rotations used to define the principal-axis coordinate systems.

\end{packed_enum}

\subsection{Force representation structure: vector valued learning versus scalar valued learning protocols}\label{sec_vector_learn}
In a high dimensional potential energy configuration space, each Cartesian force element can be regarded as a distinct scalar function of the full geometry. Forces can be learned either as separated scalar targets or vector-value targets of various block sizes. 
Their practical differences arise from the model architecture, parameter sharing, optimization behavior, and computation efficiency. In this section, we first analyze the architectural differences between scalar and vector learning with the degree of parameter sharing and its effect on prediction accuracy. We then justify our choice to decompose the full $3N$ dimensional force vector into three length $N$ vectors, each corresponding to force projections along a principal inertia axis. Each projected length $N$ vector (as obtained from the descriptors in Section \ref{sec_des}) is treated as an independent vector output target. 

\begin{figure}
    \centering
    \subfigure[Scalar training architecture]{\includegraphics[width=0.48\linewidth]{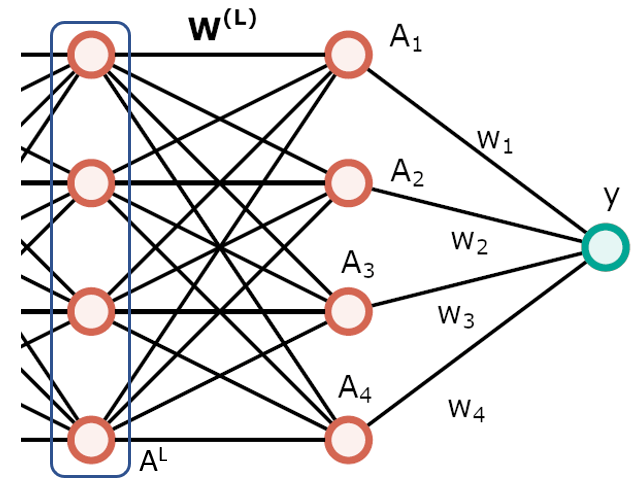}}
    \subfigure[Vector training architecture]{\includegraphics[width=0.48\linewidth]{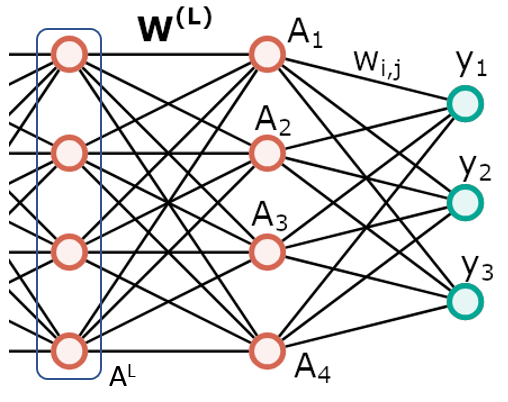}}
    \caption{A demonstration of architecture of neural network associated with (a) Eq. (\ref{eq_scalar}), 
    and (b) 
    Eq. 
    (\ref{eq_vector}). For both cases the block $A^L$ is defined by Eq. (\ref{AL-block}).}
    \label{fig_Ai}
\end{figure}

{\bf Neural networks that depend on individual components of the force vector:} The mathematical distinction between training individual force components and training the full force vector is reflected directly in the neural network architecture. Consider first a scalar-output network, corresponding to a single force component. Assuming $m$ neurons in the final hidden layer and a linear activation at the output, the prediction can be written as
\begin{align}
    y = \sum_{i=0}^{m}w_i A_i(\bar x)
    \label{eq_scalar}
\end{align}
where $A_i(\bar x)$ represents the activation of the $i$-th neuron in the last hidden layer for an input fragment geometry depicted as $\bar x$, and $w_i$ is the corresponding weights connecting to the output, as shown in Figure \ref{fig_Ai}(a). Here, $y$ is one component of the $\Delta F_{\alpha_r,r}^{ML}({\bf {\bar x}_{\alpha_r,r}})$ in Eq. (\ref{eq_graph-ML-f}). In practice, $A_i(\bar x)$ represents the complete nonlinear feature construction preceding the final hidden layer and may therefore take a much more complex functional form than a simple activation function such as
\begin{align}
    A_i({ {\bar x}}) &=  
    \sigma \left( {\bf W}_i^{(L)}  \cdot A^L({ {\bar x}}) \right),
    \label{eq_Ai}
\end{align}
where ${\bf W}_i^{(L)}$ represents the $i$-th row of the weight matrix ${\bf W}^{(L)}$ in the $L$-th hidden layer with activation function $\sigma(\cdot)$. The quantity $\left( {\bf W}_i^{(L)}  \cdot A^L({ {\bar x}}) \right)$ represents a dot product. The corresponding structure is displayed in Figure \ref{fig_Ai}(a). 
Additionally, the vector, $A^L({\bf {\bar x}})$, on the right side of Eq. (\ref{eq_Ai}) is recursively defined as 
\begin{align}
    A^L({ {\bar x}}) &=  
    \bar{\sigma} \left( {\bf W}^{(L-1)} A^{L-1}({ {\bar x}}) \right),   
    \label{AL-block}
\end{align}
where $\left( {\bf W}^{(L-1)} A^{L-1}({{\bar x}}) \right)$ now represents a matrix-vector multiplication 
and $\bar{\sigma}\left(\cdot\right)$ is a vector valued activation function. The quantity ${\bf W}^{(L-1)}$ represents a matrix with dimensions, number of neurons in hidden layer $L-1$, times the number of neurons in hidden layer $L$. Finally, $A^{1}({ {\bar x}}) \equiv { {\bar x}} $. Thus, to summarize, each force component has a separate basis from $\left\{ A_i(\bar x) \right\}$ and is represented by Eq. (\ref{eq_scalar}). Additionally each basis $A_i(\bar x) $ is obtained from the neural network represented by Eq. (\ref{eq_Ai}). 

\begin{figure}
    \centering
    \includegraphics[width=0.75\linewidth]{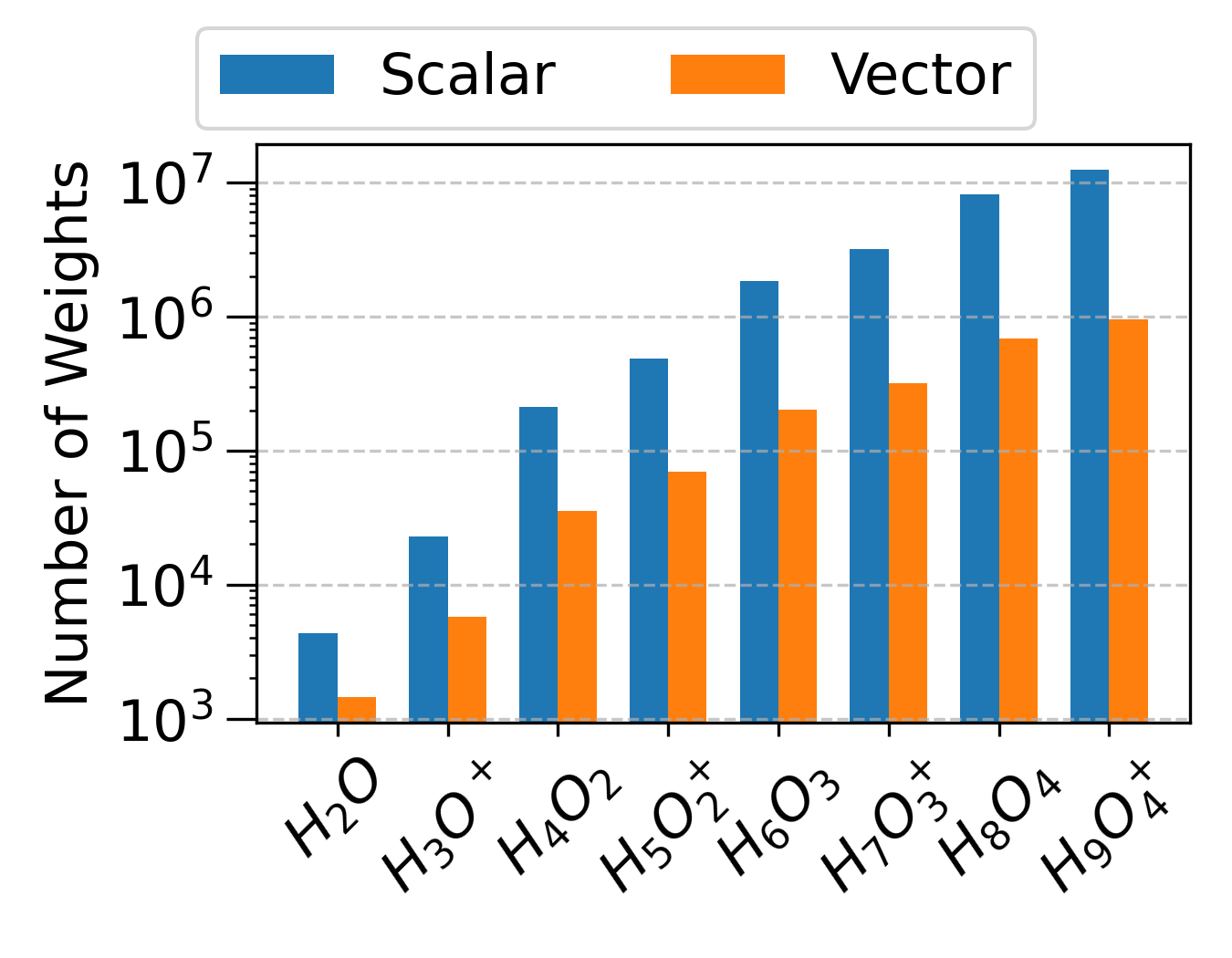}
    \caption{Number of weights (trainable parameters) for component-wise (blue) and vector-based (orange) force training strategies. Both approaches reach approximately the same level of accuracy ($\approx$0.1 milli-Hartree/Bohr) while the vector training strategies requires over one order of magnitude fewer parameters.}
    \label{fig_ele_vec}
\end{figure}

{\bf Vector valued neural networks for the force vector:} For vector-valued learning, the network architecture up to the final hidden layer remains unchanged, but the output layer now consists of $n$ neurons, one for each force component. The prediction is therefore given by
\begin{align}
    y_j = \sum_{i=0}^{m}w_{i,j}\cdot A_i(\bar x), \quad j = 1,\ldots,n,
    \label{eq_vector}.
\end{align}
where $w_{i,j}$ connects the $i$-th hidden neuron to the $j$-th force component. Figure \ref{fig_ele_vec}(b) shows the architecture for Eq. (\ref{eq_vector}).
The key difference between Eqs. (\ref{eq_scalar}) and (\ref{eq_vector}) is that vector training enforces a shared set of basis functions $\{A_i(\bar x)\}$ across all the vector components, whereas scalar training allows each force component to learn its own independent basis. 
Note that the left side of Eq. (\ref{eq_scalar}) is a specific component of the force vector, and hence each component has a different set of parameters: $\left\{ 
w_i, A_i \right\}$. This is not the case for Eq. (\ref{eq_vector}).
As a result, vector training is substantially more parameter-efficient: although the number of output weights increases linearly with the size of the vector, the dominant contribution to the total parameter count arises from the hidden-layer weights that define $\{A_i(\bar x)\}$, which are shared among all force components.

Consequently, compared to scalar training, a similar level of accuracy can typically be achieved in vector training by modestly increasing the number of hidden neurons $m$, while still using a significantly smaller total number of parameters. This parameter sharing also implicitly regularizes the model, encouraging correlated force components to be learned from common geometric features. This pattern is shown in Figure \ref{fig_ele_vec}, where we use the same fragment geometries from a solvated Zundel ($H_{13}O_{6}^+$) to train force vector model under scalar and vector forms. 
The mean absolute error of forces and total number of parameters is also listed. Clearly, as seen in Figure \ref{fig_ele_vec} for a 10$^{\text {th}}$ of the cost, the vector valued training recovers a similar error 
as the scalar valued training protocol.

Finally, we note that we construct our vector training model using a sequence of serial neural networks. In general, the $l$-th neural network in this series is trained on the cumulative error from all previous neural networks, that is, 
\begin{align}
    \min_{\left\{ {{\bf W}^{(L)}}_i \right\}} \abs{y-\sum_{i=0}^{{\cal L} -1}y^p_{l}\left({\bar x},{\left\{ {{\bf W}^{(L)}} \right\}_l}\right)}^2.
    \label{eq_nn_array}
\end{align}
where $y^p_i$ is the prediction from the $i$-th neural network and $\left\{ \left\{ {{\bf W}^{(L)}} \right\}_i \right\}$ are the weight matrices for the $i$-th neural network. In this work, we use $l=2$ for each force vector component training.

\subsection{Summary of energy and force neural network approximations}\label{sec_sum}
Based on the discussion above, we may combine Eqs. \ref{eq_graph-ML}, \ref{eq_scalar} and \ref{eq_nn_array} to write an overall potential energy prediction expression as a scalar learning formalism as
\begin{align}
\Delta E_{\alpha_r,r}^{ML}({\bf {\bar x}_{\alpha_r,r}}) = \sum_{l=0}^{\cal L} \sum_{i=0}^m {w}_{\alpha_r, r}^{E,l,i}
A_{\alpha_r, r}^{E,l,i}({\bf {\bar x}_{\alpha_r,r})}.
\end{align}
where our scalar weights and associated basis functions
are $\left\{ {w}_{\alpha_r, r}^{E,l,i};A_{\alpha_r, r}^{E,l,i}({\bf {\bar x}_{\alpha_r,r})}\right\}$. 
Therefore, 
\begin{align}
\Delta E^{ML}({\bf {\bar x}})=
\sum_{r=0}^{\cal R}  { \sum_{\alpha_r  \in {\bf V}_r}{\cal M}_{\alpha_r, r}^{\cal R} \; 
\sum_{l=0}^{\cal L} \sum_{i=0}^m {w}_{\alpha_r, r}^{E,l,i}\ A_{\alpha_r, r}^{E,l,i}({\bf \bar x}_{\alpha_r,r})}.
\label{eq_graph-ML-final}
\end{align}
The neural network weights ${w}_{\alpha_r, r}^{E,l,i}$ contains superscript $E$ which denotes that these parameters are for an energy model specifically trained for fragment $\alpha_r$ at rank $r$. The superscript $l$ denotes it is the $l$-th serial neural network in the array from Eq. (\ref{eq_nn_array}) and $i$ represents that the weight is associated with the $i$-th abstract basis function $A_{\alpha_r, r}^{E,l,i}({\bf \bar x}_{\alpha_r,r})$ from Eqs. (\ref{eq_scalar}) and (\ref{eq_Ai}). 

This function itself is directly differentiable with respect to input $\bf{\bar x}$, but the accuracy of derivatives in general depends heavily on the neighboring data points along each dimension, and not on the data points themselves. This is expected to substantially increase the training costs 
in higher dimensions. 

To overcome this computational bottleneck, we directly construct neural networks for forces as discussed here. Hence, for the 
vector model
\begin{align}
\Delta F_{j; \alpha_r,r}^{ML}({\bf {\bar x}_{\alpha_r,r}}) = \sum_{l=0}^{\cal L} \sum_{i=0}^m {w}_{\alpha_r, r}^{F,l,i,j}\ A_{\alpha_r, r}^{F,l,i}({\bf \bar x}_{\alpha_r,r})
\end{align}
and
\begin{align}
\Delta F_j^{ML}({\bf {\bar x}}) =&
\sum_{r=0}^{\cal R}  \sum_{\alpha_r  \in {\bf V}_r}{\cal M}_{\alpha_r, r}^{\cal R} \; 
\nonumber \\ & \sum_{l=0}^{\cal L} \sum_{i=0}^m {w}_{\alpha_r, r}^{F,l,i,j}\ A_{\alpha_r, r}^{F,l,i}({\bf \bar x}_{\alpha_r,r})\frac{\partial ({\bf {\bar x}}_{\alpha_r,r})}{\partial ({\bf {\bar x}}^j)}.
\label{eq_graph-ML-F-final}
\end{align}
Here the neural network weights ${w}_{\alpha_r, r}^{F,l,i,j}$ and generalized activations, $A_{\alpha_r, r}^{F,l,i}({\bf \bar x})$ 
are from Eq. 
(\ref{eq_vector}) but contain the superscript $F$. This denotes that these parameters are for the vector force model and different from the parameters in Eq. (\ref{eq_graph-ML-final}), since these are specifically trained for fragment $\alpha_r$ forces. 

The key idea here is of course the independent treatment of energy and forces (Eqs. (\ref{eq_nn_array}) and (\ref{eq_graph-ML-F-final})) which reduces the training needed for forces. As we will see, this results in 10\% learning from AIMD data, with use of the tesselation algorithm as will be discussed in the result section. For all calculations below, the vector model is used for forces and the scalar model for energies.

Although atomic forces are naturally represented as vectors in $3N$-dimensional space, their statistical distribution is generally anisotropic and depends strongly on the direction of molecular motion. For example, force components projected along stiff bond directions often exhibit substantially different magnitudes and variances than those associated with softer modes, such as hydrogen-bond, bond-angle bending, or librational motions. Consequently, treating all force components as elements of a single undifferentiated target vector mixes quantities with heterogeneous statistical scales, increasing the complexity of the learning problem. To address this issue, we exploit the principal-axis coordinate system introduced in Section~\ref{sec_des} to obtain fragment descriptors. Specifically, the atomic forces are projected onto the three fragment-fixed principal axes (Section~\ref{sec_des}), and the resulting projections are organized into three separate length-$N$ vectors, each containing the force components along one principal-axis direction, as described in Section~\ref{sec_des}. Each block therefore represents force components associated with a single, physically meaningful direction determined by the instantaneous mass distribution. This directionally aware decomposition of forces enables the associated force neural networks to model the distinct statistical characteristics of force components along each principal axis independently. In effect, the transformation acts as a statistical standardization of the force targets, improving both the conditioning of the learning problem and the interpretability of the resulting models.

\section{Computational experiments that probe the accuracy of ML-generated nuclear gradients}\label{sec_result}
The solvated Zundel cation plays a central role in understanding proton transfer in aqueous environments because it captures the essential structures that govern proton migration in water clusters. In particular, protonated water clusters dynamically inter-convert between configurations resembling the Zundel (${H_5O_2^+}$) and Eigen (${H_9O_4^+}$) species, which represent the shared proton and symmetrically localized hydronium description of the excess proton\cite{zundel2012hydration,cptuckerman3,h+oh-solv,schmittvothprotontransport1999,schmitt1998multistate}. The interconversion between these species provides fundamental insight into proton transport mechanisms across a broad range of biological, atmospheric, material and condensed phase systems\cite{HDMeyer-Zundel-1,protonwire1,atmosph-clusters1,atmosph-clusters2,pomesroux2,protonwire3,protonwire4,teeter,bio-clusters2,lipscombpeek,turowlett,Scott-proj,Harry-weighted-graphs}.
Accurate description of these systems is challenging due to (a) critical quantum nuclear effects arising from the shared proton dynamics, (b) coupling of donor acceptor vibrational modes with proton transfer modes across multiple dimensions aa manifested from the Grotthuss mechanism for proton transfer\cite{Agmon:95} and (c) the highly polarized electronic structure arising from the delocalized nature of the shared protons in the system. Computationally prohibitive post-Hartree Fock methods\cite{H5O2+XCcomp,Schaefer:94,DFT-vDW-Michaelides,Truhlar-functional-review,yang-dft-review} are generally required to achieve benchmark quality for energy and force calculation. 
However, given the demand for computing energies and forces at every nuclear dynamics step, AIMD is drastically limited in its ability to lend support for the study of such complex problems. 

With the goal of generating ML-aided post-Hartree-Fock AIMD trajectories, here we ask if the nuclear gradients obtained using the algorithms described above provide accurate forces. We also evaluate the corresponding computational cost reduction. We begin by considering an AIMD trajectory generated previously\cite{fragAIMD-CC,frag-ML-Xiao,Xiao-LLM} as a baseline for comparing nuclear forces. This trajectory consists of 9326 frames of solvated Zundel geometries spanning a total of 1.86 pico-seconds. Each geometry is treated via graph-theoretic fragmentation using Eq. (\ref{eq_graph-main}) with ${\cal R}=1$. The high level electronic structure theory is CCSD and the low level theory is B3LYP, both using the 6-31+g(d, p) basis set. However, it must be noted that this level of AIMD is not necessary as a starting point and simpler sampling strategies will suffice as we explain below. These aspects will be probed to a greater extent in future publications. 

In this specific trajectory, we treat each oxygen as the center of a node and group all hydrogens within 1.4{\AA} of that oxygen into the node. The edges in the graph are created dynamically by defining a minimum oxygen-oxygen distance for each oxygen ($D_i$). A maximum of three nodes within $1.1*D_i$ are connected to the $i$-th node. A range of structures is obtained in this manner to evaluate the accuracy of our the forces obtained using the protocols introduced above. 

\begin{figure}
    \centering
\subfigure[oxygen-oxygen distance distribution]{\includegraphics[width=0.49\linewidth]{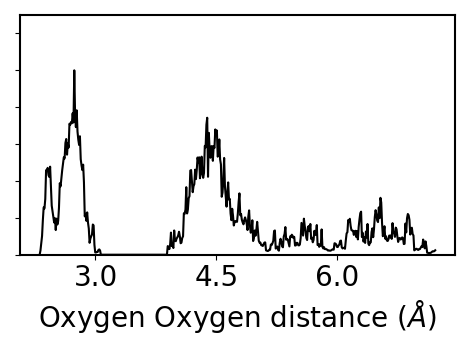}}
\subfigure[oxygen-oxygen-oxygen angle distribution in $H_6O_3$]{\includegraphics[width=0.49\linewidth]{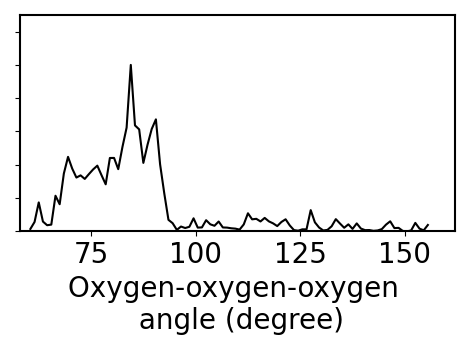}}
\subfigure[oxygen-oxygen-oxygen angle distribution in $H_7O_3^+$]{\includegraphics[width=0.49\linewidth]{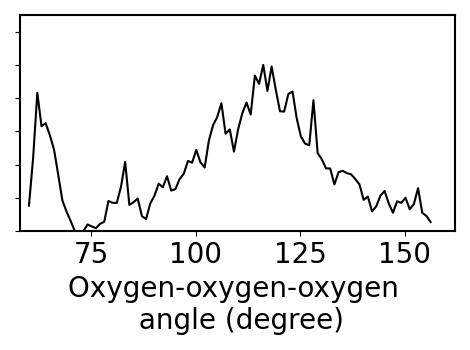}}
    \caption{Distribution functions that determine the range of structures used for constructing ML models.}
    \label{fig:RDF}
\end{figure}
To evaluate the structural diversity of molecular and fragment configurations obtained from this starting trajectory, we present a few critical distribution functions in Figure \ref{fig:RDF}. Figure \ref{fig:RDF}(a) shows the OO distribution which affects the water dimer fragments and involves primary and secondary solvation shell interactions. Figures \ref{fig:RDF}(b) and \ref{fig:RDF}(c) provide similar structural assessments for secondary structures involving all water trimers. 
As shown in Figure \ref{fig:RDF}(a), the 
wide range of oxygen-oxygen distances 
accounts for both classical hydrogen-bonded water dimers and non-bonded spatial combinations of water molecules residing in the first and second solvation shells.

For the trimer fragments, we show the largest oxygen-oxygen-oxygen angle within each distinct fragment in Figures \ref{fig:RDF}(b) and \ref{fig:RDF}(c). The neutral and protonated species exhibit remarkably different distributions because the excess proton remains consistently localized within the central Zundel-like configuration. As a result, the neutral fragments tend to be non-bonded trimers that exhibit smaller angles, whereas the protonated trimers tend to be hydrogen bonded changes and display a propensity for larger angles. 
Additionally, the protonated trimers also  show a significant population of non-bonded chains, as exhibited by the distribution at lower angles. 

Given the large diversity for structures 
with significant non-bonded interactions involved, these data sets present a critical challenge to ML, but as we will see, using protocols established here we do 
construct suitable approximations to the full system forces accurately and produce ML-based AIMD trajectories that agree well with regular AIMD trajectories. 

\subsection{Geometric tesselation of the dataset to obtain representative training geometries for suitable energy and force extrapolations}
\label{sec_sampling}

Given the diverse nature of structures presented above, it is key that we develop a sampling strategy that appropriately covers all regions of the configurational space and lens naturally to ML training. The desirable features of such a sampling strategy are as follows: (a) The sampling strategy must be general and work for arbitrary dimensional fragments. (b) It must have suitable controls for fine and coarse-graining of the fragment data space as needed based on the level of computational power. (c) It must be adaptive, that is the level of fine-graining can be changed as dictated by the needed accuracy. 

Based on these criteria, 
we employ a k-means\cite{kmeans,frag-ML-Xiao} based clustering strategy to select 
representative geometries on the potential energy surface space of fragments for training our force neural networks. By selecting the cluster centroids from such a multi-dimensional tessellation algorithm\cite{macqueen1967some,Bowyer_Dirichlet,Voronoi-1,cgal-book}, 
we 
balance the geometric variations across all degrees of freedom. This creates a shared and tunable training set that accommodates the distinct training requirement of all concatenated force elements simultaneously.  

The standard k-means algorithm\cite{kmeans} is a high-dimensional space tessellation algorithm. It divides the fragment data space into $k$ mutually exclusive regions called clusters represented here as $\{C_j\}$,  and is known to provide compact representations for large datasets\cite{macqueen1967some}. Each cluster, $C_j$, in the multi-dimensional fragment potential energy space, is represented by its centroid 
$\langle \bf{\bar x}_j \rangle$, with all data points being assigned to the closest centroid to define a specific 
cluster. In the k-means algorithm, centroid positions are iteratively updated as the arithmetic mean of all data points inside the cluster 
until these centroid positions converge. 
If we use $\left\{ \bf{\bar x}_i^j \right\}$ to represent the 
the set of structures inside the
$j$-th cluster, $C_j$, for a given fragment, 
the clustering algorithm finds a preset number of clusters ($k$, below) and their centroid positions by minimizing the cost function
\begin{align}
    \min_{\langle \bar{\bf {x}}_i \rangle} \sum_{j=1}^{k} \sum_{\bf{\bar x}_i^j \in C_j} |{\bf{\bar x}_i^j-\langle {\bf{\bar x}}_j \rangle }|^2.
    \label{kmeans-cost-fn}
\end{align}
For our purpose, as discussed in Section \ref{sec_des}, the quantities ${\bf{\bar x}_i^j}$ and $\langle {\bf{\bar x}}_j \rangle$ would represent symmetry preserving interatomic distance descriptors that are projected onto the moment of inertia axes of the fragments. 

However, it must be emphasized that here we use the mini-batch variation of the k-means algorithm\cite{mini-batch-kmeans,kmeans} above, which is a variation of this standard k-means algorithm and uses a random subset of data (known as a batch) from the original dataset (the AIMD trajectory above) to update the centroid positions during each iteration. The batch feature significantly reduces the computational overhead in finding the centroids. 
The geometric properties of compact data representations produced by k-means \cite{macqueen1967some} ensure that each centroid corresponds to a localized and relatively uniform region of the configuration space, thus providing balanced coverage of structural variations across all degrees of freedom.
Once the centroids and the corresponding representative data are identified, they are used to train fragment level network  models to represented nuclear gradients as described above. Critically, the algorithm here also provides us with a fine-tuning knob, where the level of training needed can be changed based on the desired coverage of the potential energy space. This is a key feature and as we will find, this feature can be used to improve accuracy of the network models as needed. 

\subsection{Accuracy of fragment forces}\label{sec_frag}

\begin{table}[h!]
\centering
\begin{tabular*}{\columnwidth}{@{\extracolsep{\fill}}lccc}
\hline
Fragments& \begin{tabular}{@{}c@{}}Total\\ ($N_{Config}$)\end{tabular}  & \begin{tabular}{@{}c@{}}Training-data\\ size\end{tabular} &
\begin{tabular}{@{}c@{}}Force error Eq.(\ref{Force-Error-NT})\\in mEh/Bohr\end{tabular} \\
 
 \hline  
$H_2O$ &	38910&3891&	0.000\\ 
$H_3O^+$ &	17046&1704&	0.002\\
$H_4O_2$ &	62380&6238&0.008\\
$H_5O_2^+$ &	77510&7751&0.012\\
$H_6O_3$ &	46940&4694&0.012\\

$H_7O_3^+$ &	139580&13958&	0.017\\ 

$H_8O_4$ &	15750&1575&0.011\\ 
$H_9O_4^+$ &	124140&12414&0.020\\ 
\hline
\end{tabular*}
\caption{The number of fragments generated from the reference trajectory geometries for fully connected graph at rank ${\cal R}=3$ and their corresponding force error. Also see Figures \ref{fig_frag_force_error} and \ref{Error-distr}.}
\label{tab_force_error}
\end{table}

In Ref. \onlinecite{frag-ML-Xiao}, we used a fixed ratio of 10\% of fragments obtained from k-means processing of the 
aforementioned reference AIMD trajectory to train energy models for the potential energy surfaces for fragments within the solvated Zundel cation. Although the trajectory is generated from the dynamic edge cutoff at rank 1, we take all the geometries and apply fully connected graph representations to yield a fragment database for model training. (As noted earlier the method of generating the initial AIMD data is independent of the training algorithm is constructed.) The number of fragments and the training-set sizes for each fragment type are shown in Table \ref{tab_force_error}. The 10\% data was obtained using the k-means method described above. Here we prepare a set of pretrained force models following the 
procedure in Section \ref{sec_force}, 
using 10\% fragment geometries initially to evaluate the accuracy of force neural network models using descriptors introduced in Section \ref{sec_des}. To probe this accuracy, 
for different types of fragments, we compute the average L2-norm of error vector per dimension for each fragment type and use these to obtain the cumulative error in forces for a given fragments as, 


\begin{figure}
{\includegraphics[width=0.9\linewidth]{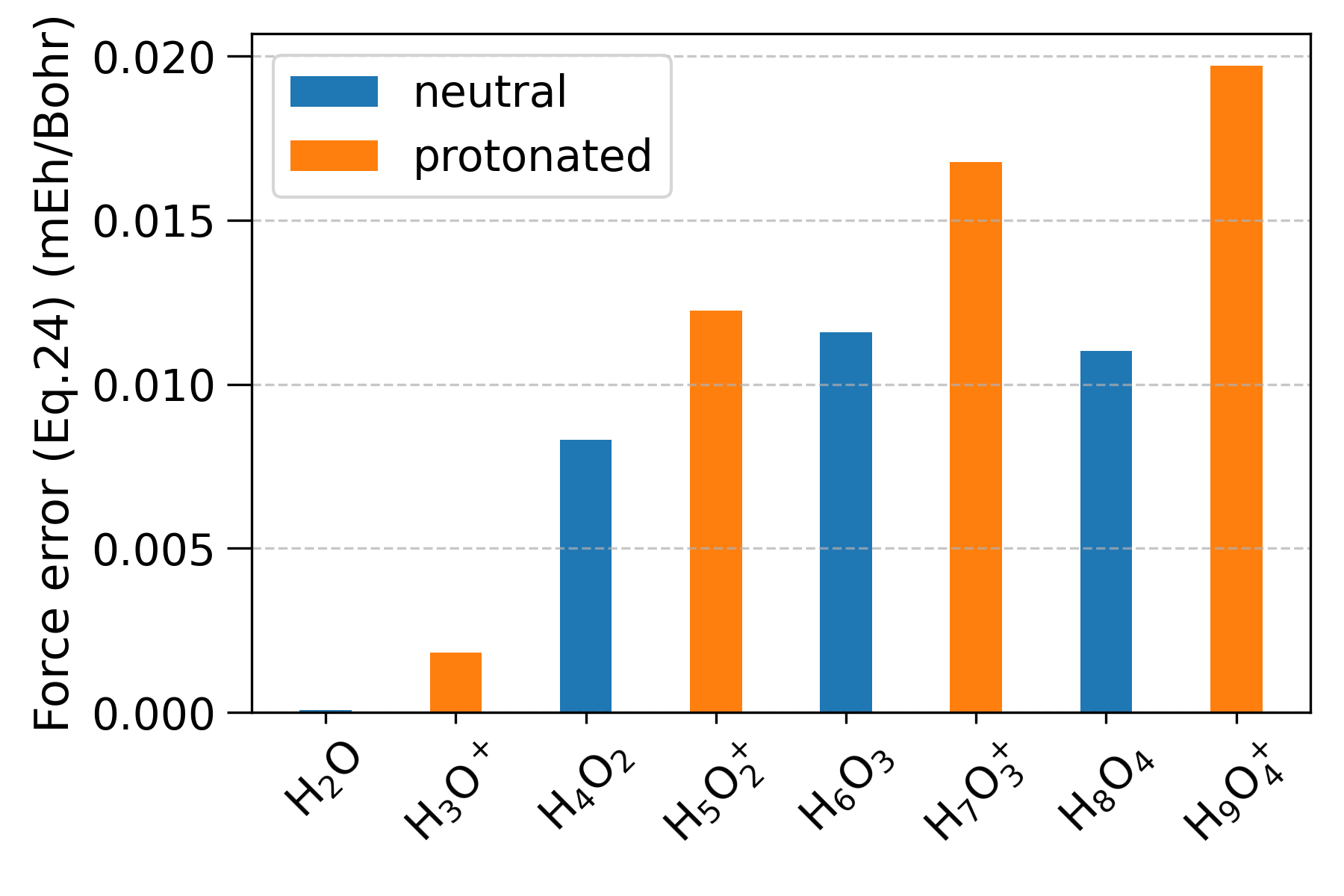}}
    \caption{Error as shown in Eq. (\ref{Force-Error-NT}). This indicates the net additional force per dimension due to ML error. }
    \label{fig_frag_force_error}
\end{figure}

\begin{figure}
\subfigure[H$_4$O$_2$]{\includegraphics[width=0.23\textwidth]{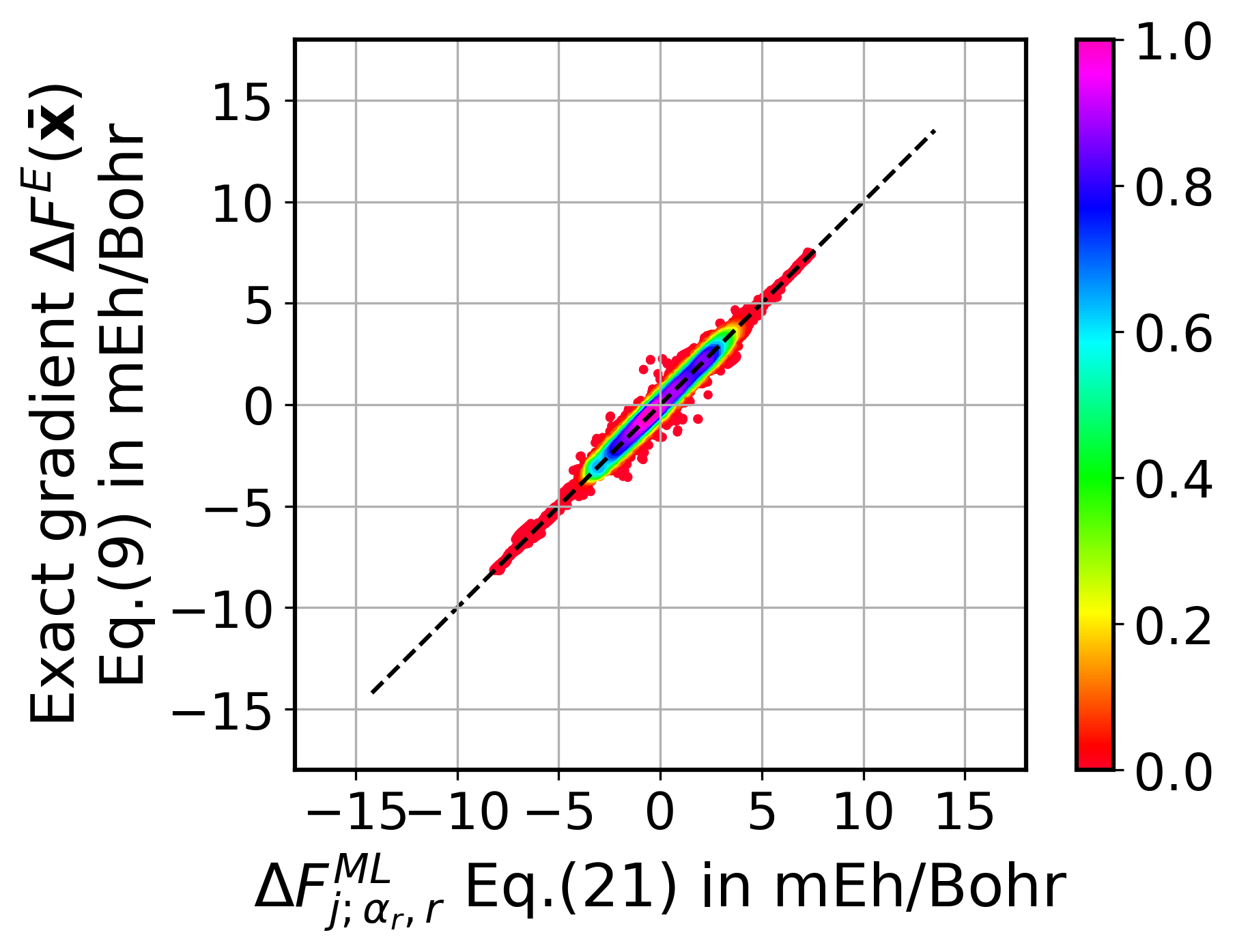}}
\subfigure[H$_5$O$_2^+$]{\includegraphics[width=0.23\textwidth]{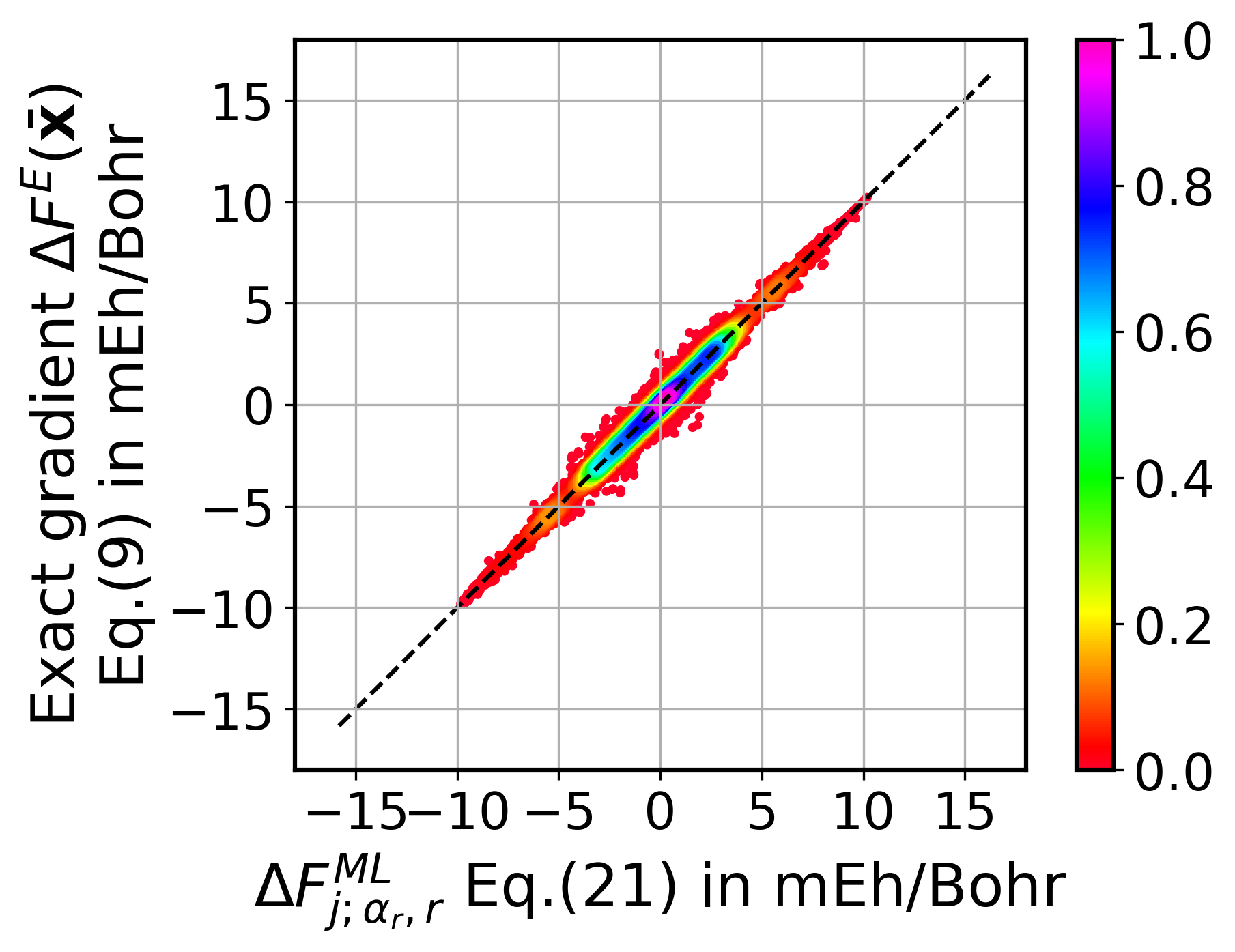}}
\subfigure[H$_6$O$_3$]{\includegraphics[width=0.23\textwidth]{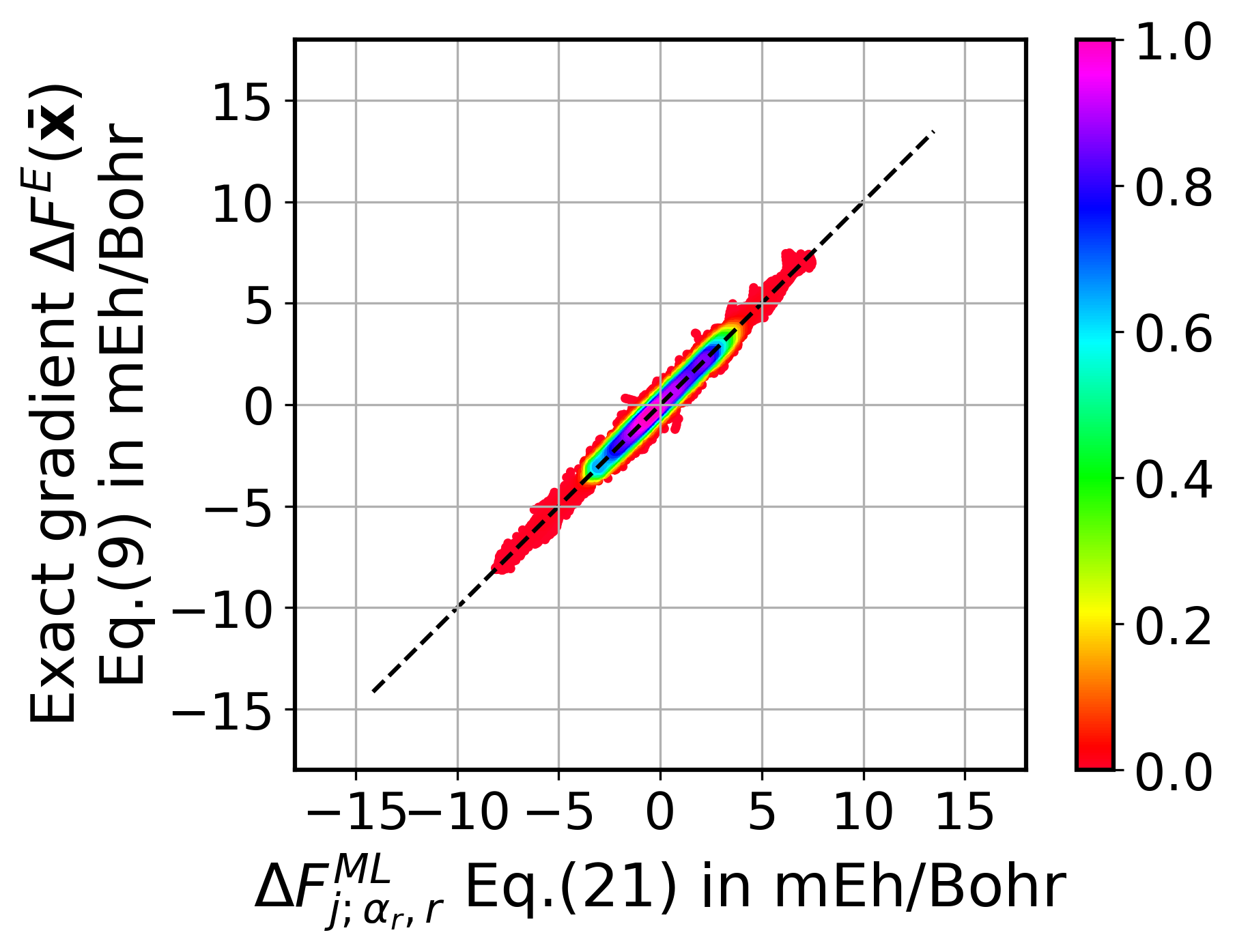}}
\subfigure[H$_7$O$_3^+$]{\includegraphics[width=0.23\textwidth]{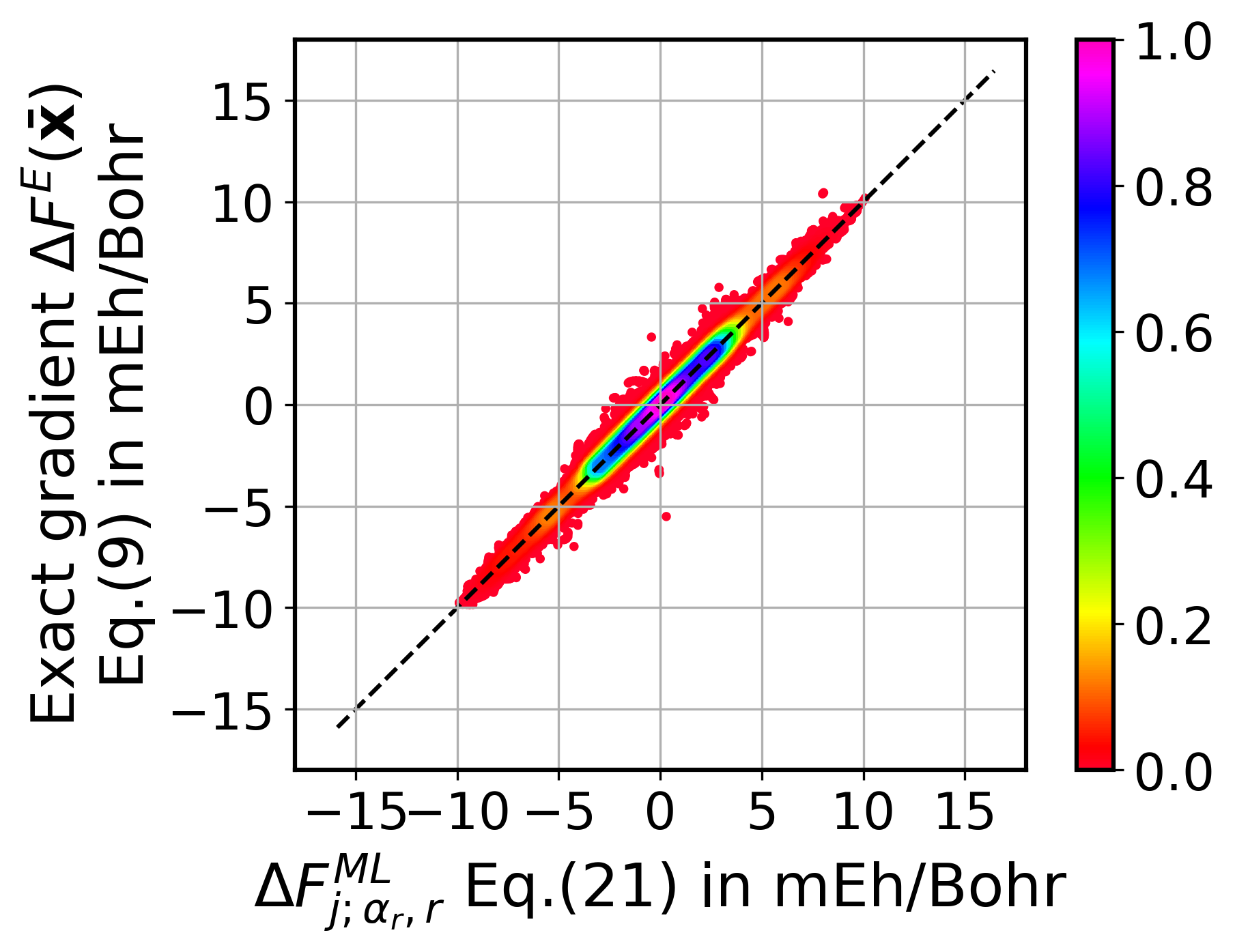}}
\subfigure[H$_8$O$_4$]{\includegraphics[width=0.23\textwidth]{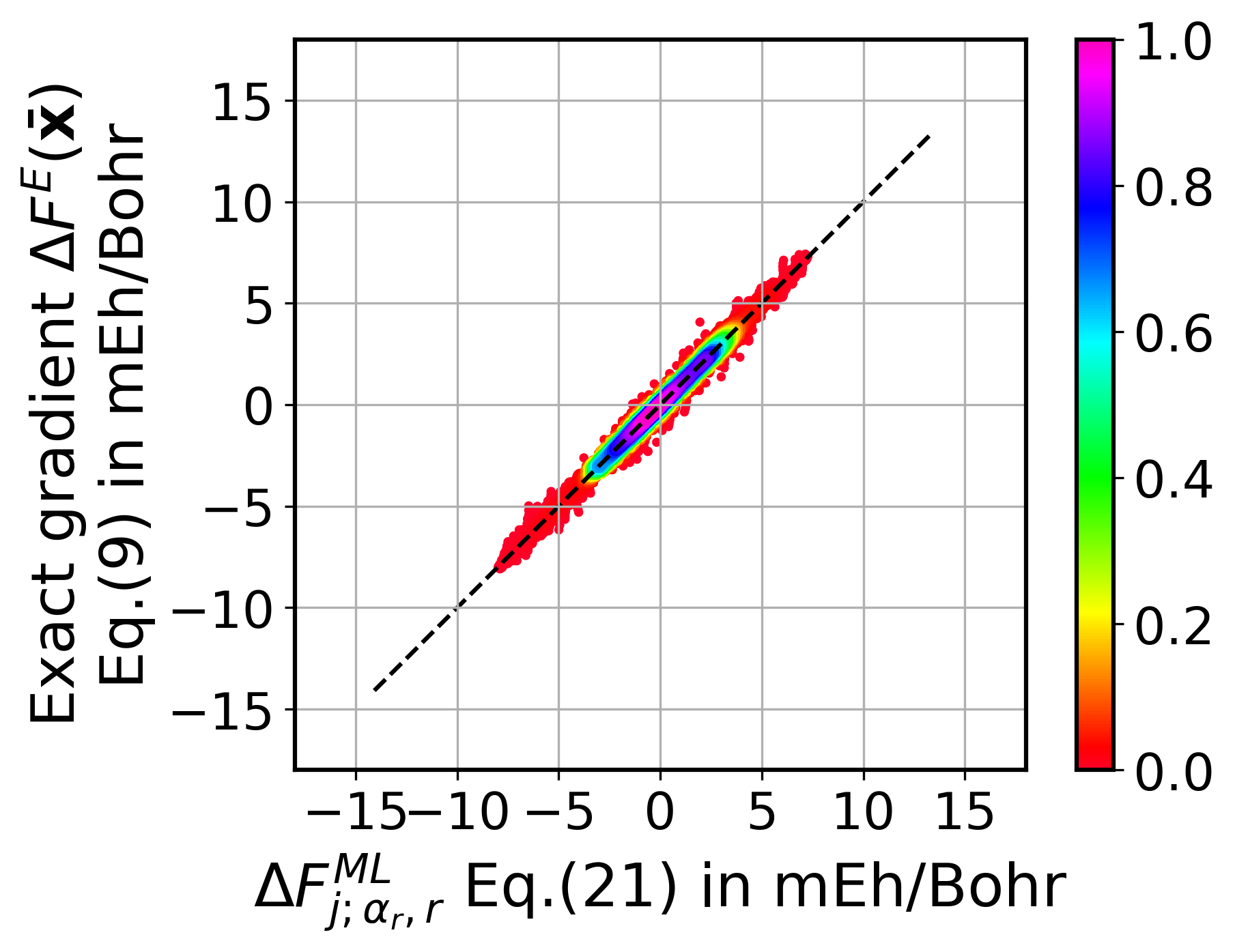}}
\subfigure[H$_9$O$_4^+$]{\includegraphics[width=0.23\textwidth]{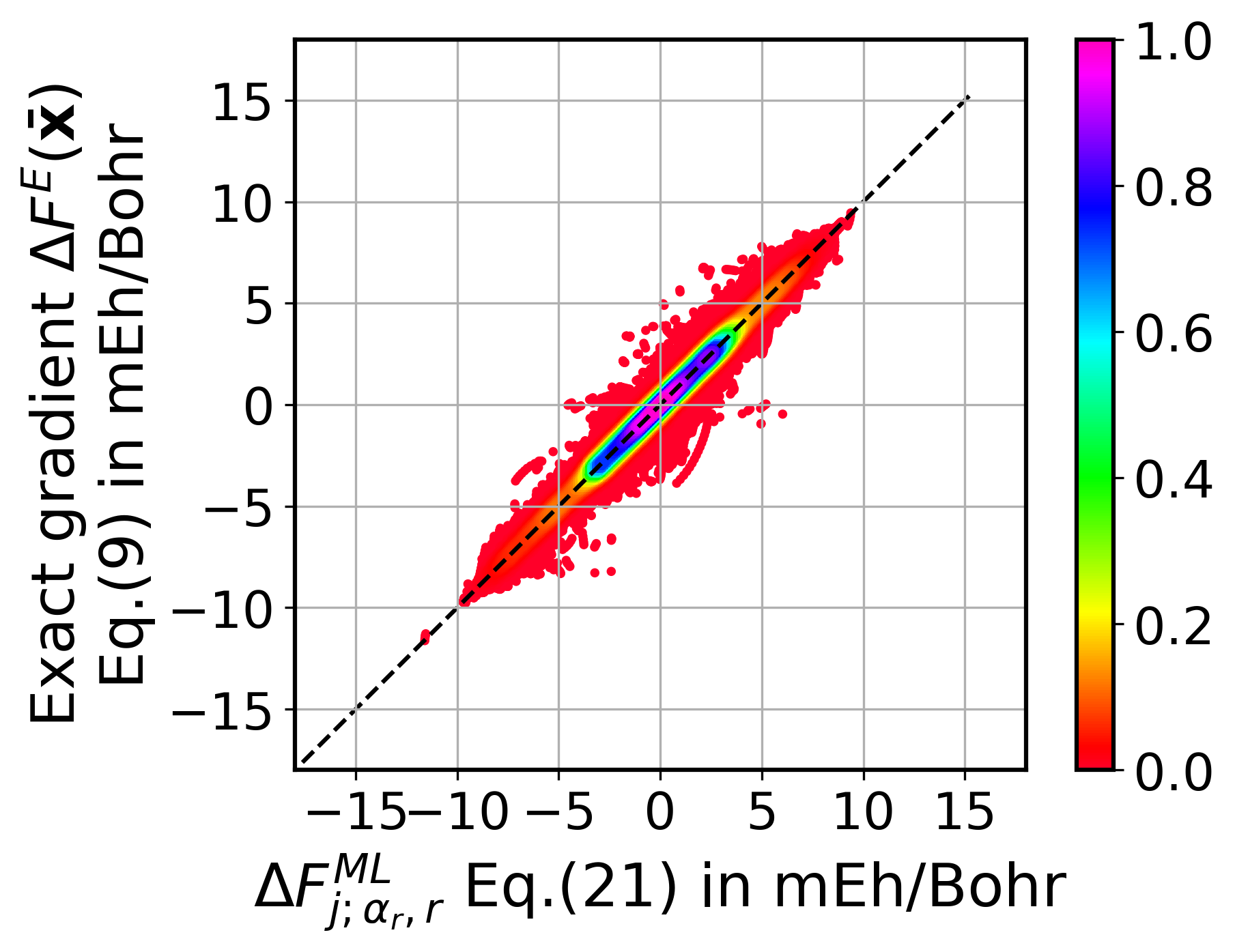}}

\caption{\label{Error-distr} Distribution of fragment force errors. Given the range of fragment structures available for each fragment as seen in Figure \ref{fig:RDF}, the ML force models provide good estimates for the exact coupled cluster values.. }
\end{figure}

\begin{align}
    \frac{1}{N_{Config}*N_{Deg}} \sum_s^{N_{Config}} \left\| \Delta F^{ML,s}_{\alpha_r,r} - \Delta F^{E,s}_{\alpha_r,r} \right\|_2
    \label{Force-Error-NT}
\end{align}
where $\Delta F^{ML,s}_{\alpha_r,r}$ is the ML prediction needed in Eq. (\ref{eq_graph-ML-f}), and $\Delta F^E_{\alpha_r,r}$ is the corresponding exact value, each of these for a specific geometry $s$ from the dataset. The quantity $N_{Config}$ is the number of fragments used to compute the force error, and $N_{Deg}$ is the number of degrees of freedom in the specific fragment (which is three times the number of atoms in each fragment). Thus, Eq. (\ref{Force-Error-NT}) represents the configurationally-averaged $L^2$-norm of the force error vector per degree of freedom and is hence the additional force on the nuclear degrees of freedom as a result of the ML prediction, which in turn will affect the quality of molecular dynamics simulations obtained from such ML force fields, and optimization calculations. In Figure \ref{fig_frag_force_error} and Table \ref{tab_force_error} we show the error 
as given by Eq. (\ref{Force-Error-NT}). As can be seen these errors are $< 0.02$-milli-Hartree/Bohr range and are well within the RMS cartesian force cut-off in optimization calculations performed during standard electronic structure calcualtions\cite{g16}. For reference, in standard electronic structure packages, a geometry optimization calculation is declared as converged when the RMS forces are of the order of 100$^{\text{th}}$-milli-Hartree/Bohr (or 10-micro-Hartrees/Bohr) and the errors in Figure \ref{fig_frag_force_error} are well within this range. This suggests that the fragment neural networks are accurate despite the wide variation in structures in our dataset. 

In Figure \ref{Error-distr}, we  additionally provide the distribution of component-wise errors for each fragment type. 
Clearly the machine learning model for forces provides an overall good representation of forces for geometries that are connected and those that are weakly bound. A detailed illustration of model accuracy for bonded and non bonded species can be found in Appendix \ref{sec_division}. By comparing the error distribution between the protonated and neutral species at each rank, a clear pattern can be seen that protonated species are generally associated with higher error. This is because the protonated species has larger geometrical variations as shown by the wider distribution of oxygen-oxygen-oxygen angles for $H_7O_3^+$ in Figure \ref{fig:RDF}(c).

\subsection{Full system force accuracy from graph-based fragmentation without machine learning}\label{sec_result_full}

\begin{figure}
        \subfigure[Dynamic edge cutoff]{\includegraphics[width=0.49\linewidth]{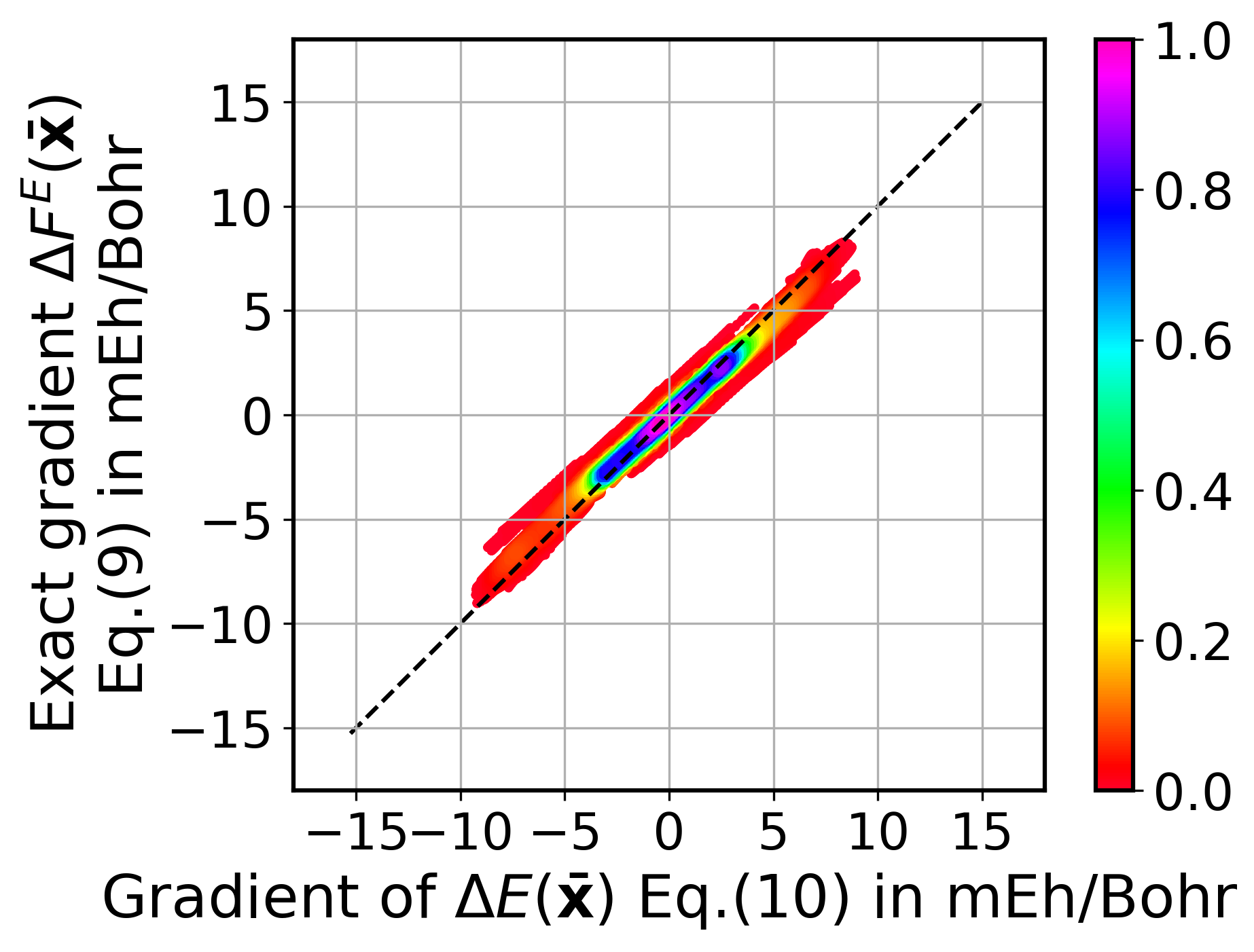}}
        \subfigure[${\cal R}=1$]{\includegraphics[width=0.49\linewidth]{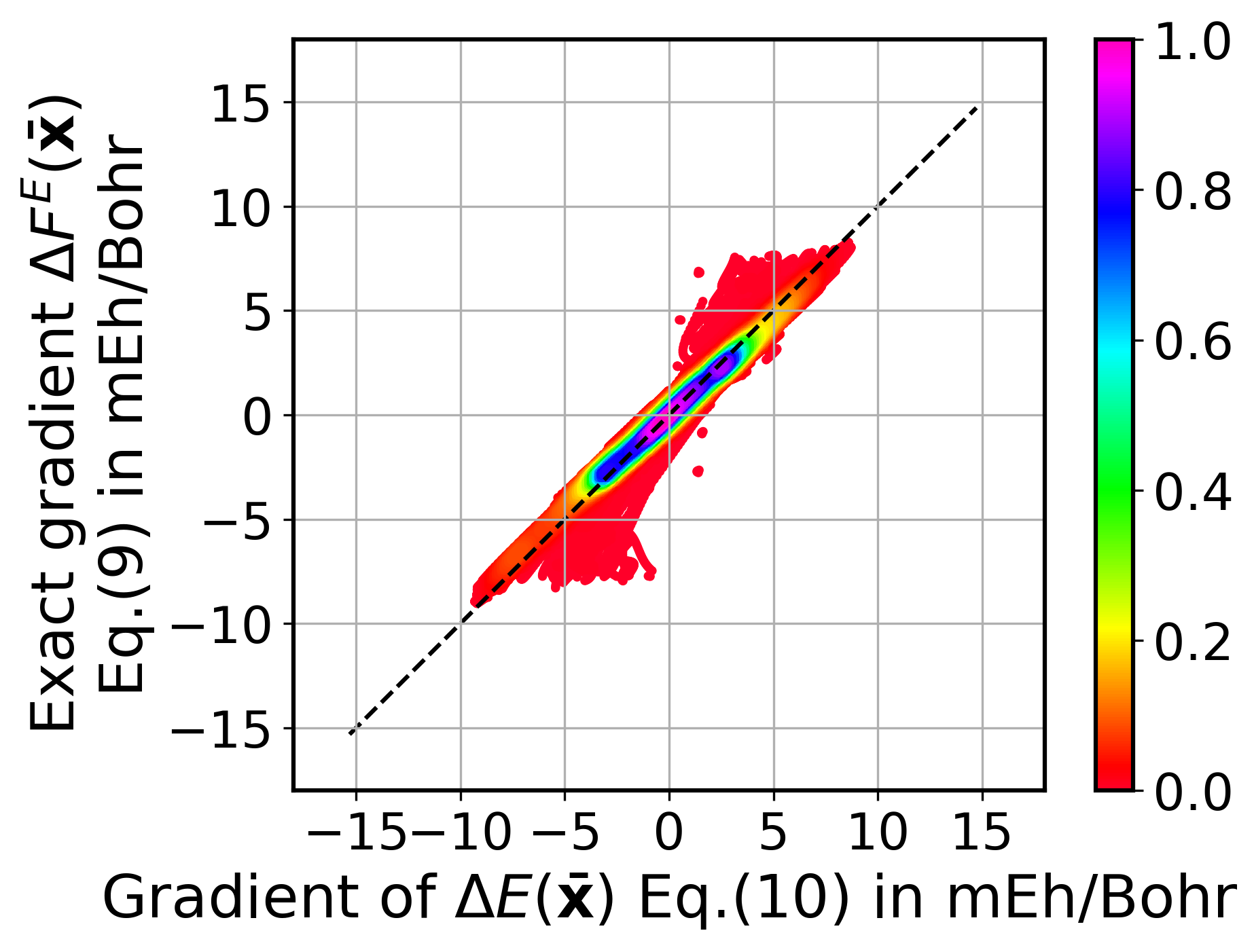}}
        \subfigure[${\cal R}=2$]{\includegraphics[width=0.49\linewidth]{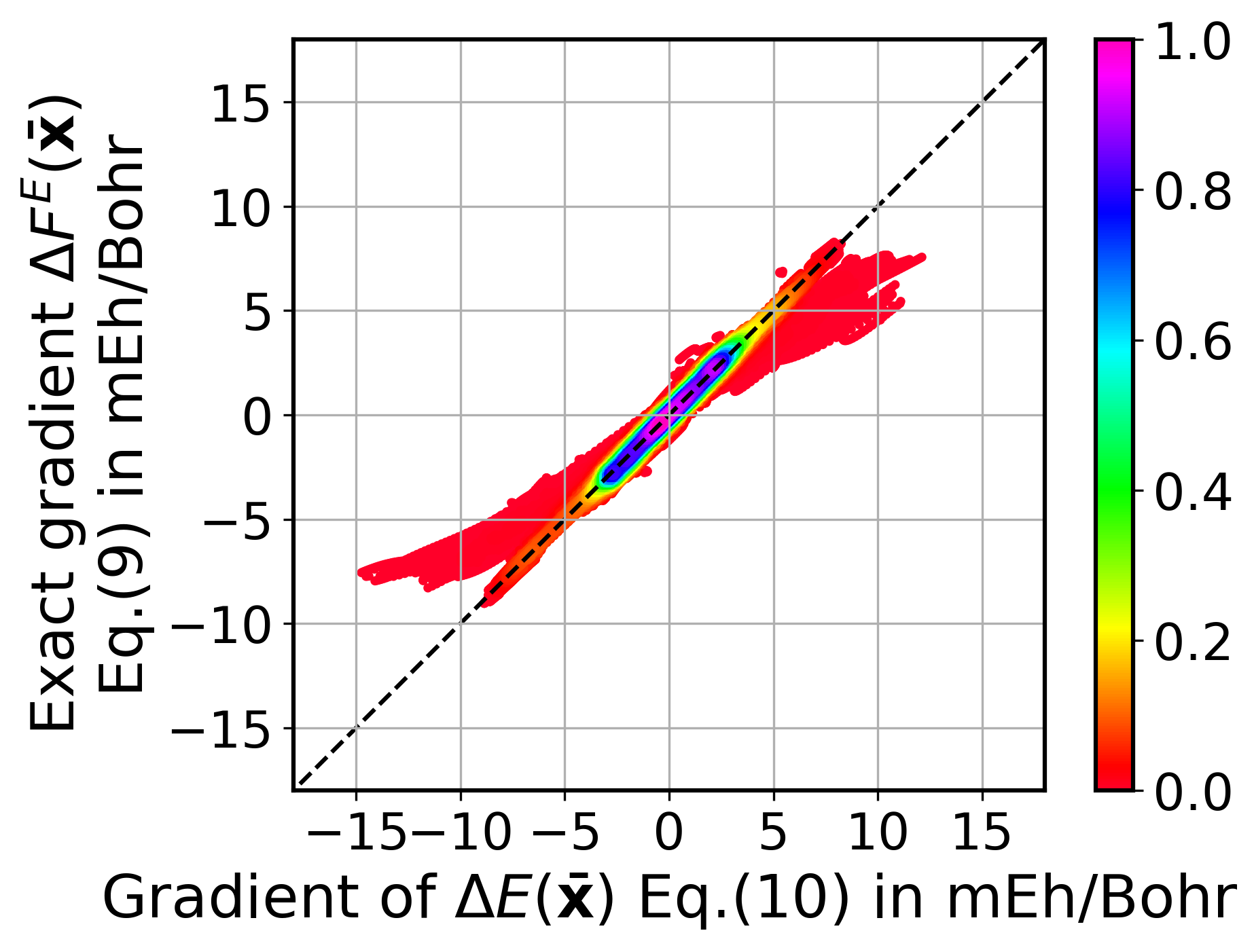}}
        \subfigure[${\cal R}=3$]{\includegraphics[width=0.49\linewidth]{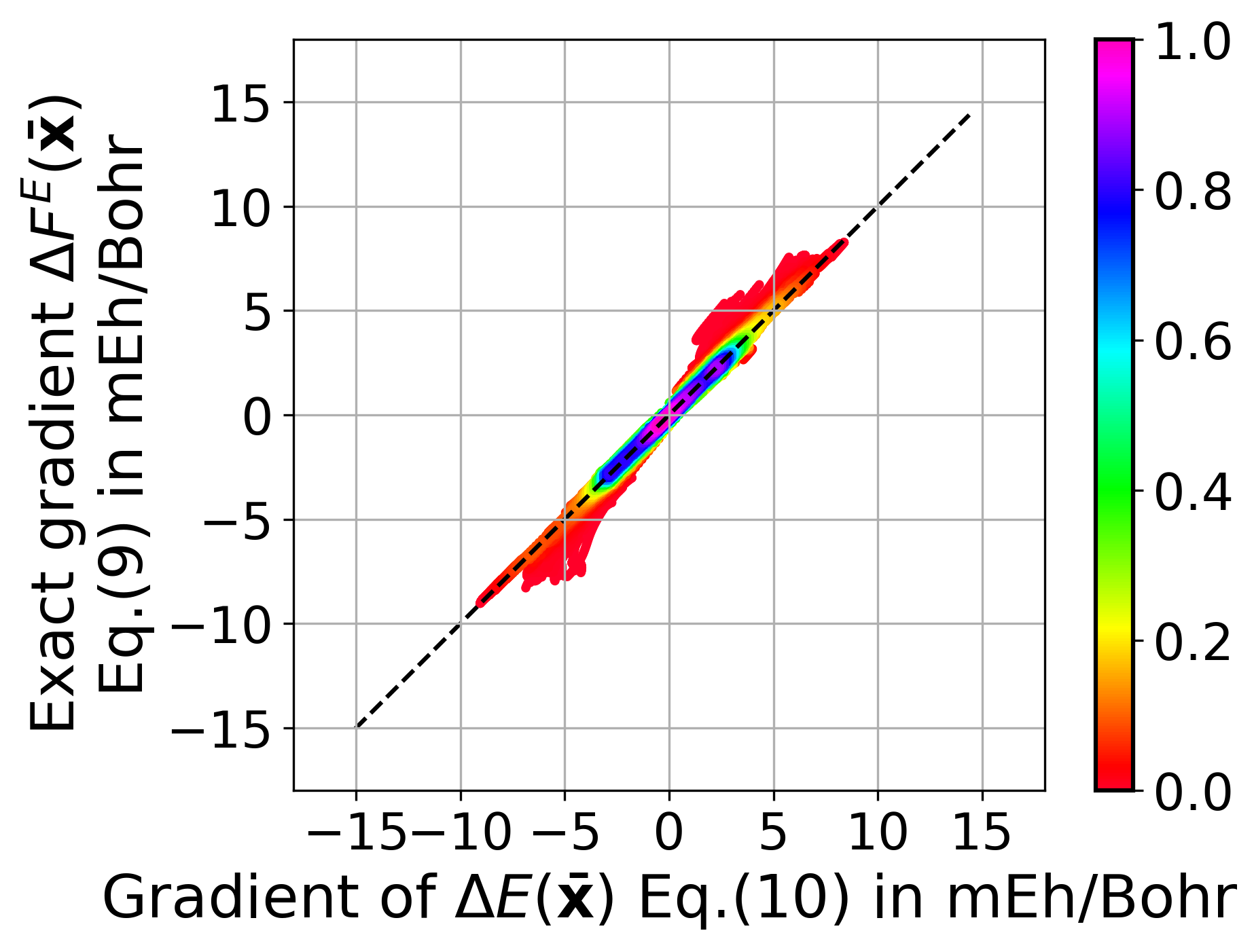}}
    \caption{Full system force component and graph force component comparison at different maximum rank cutoff for fully connected graphs.}
    \label{fig_graph_force_comp}
\end{figure}

We establish a baseline for molecular fragmentation here before evaluating ML accuracy. We evaluate the accuracy of $\frac{\partial \Delta E({\bf {\bar x}})}{\partial ({\bf {\bar x}})} $, in Eq. (\ref{eq_graph-f}) relative to the reference full system force differences evaluated at the CCSD and B3LYP levels of theory for various values of ${\cal R}$ and fragmentation protocols. 
Figure \ref{fig_graph_force_comp} shows the force error distribution for all components. 
The horizontal axis in Figure \ref{fig_graph_force_comp} represents the the forces obtained using the graph formalism in Eq. (\ref{eq_graph-f}), whereas the vertical axis presents the exact value. 
In addition to standard fully connected graph, we also include a rank label `Dynamic edge cutoff' which indicates a dynamic edge distance scheme introduced in Ref. \onlinecite{fragAIMD-CC}. As noted above, in this dynamic framework, each oxygen is treated as the center of a node with the node comprising the group of all hydrogen atoms that are within 1.4{\AA} of that specific oxygen atom. 
We dynamically construct the graph's edges using a localized threshold: for each oxygen atom $i$ with a nearest-neighbor oxygen-oxygen distance $D_i$, an edge is created with all neighboring oxygen atoms that are within $1.1 \times D_i$.

\begin{table}[h!]
\centering
\begin{tabular*}{\columnwidth}{@{\extracolsep{\fill}}lcccc}
\hline \hline
Rank ($\cal{R}$) &
\begin{tabular}{@{}c@{}}$\frac{\partial \Delta E({\bf {\bar x}})}{\partial ({\bf {\bar x}})}$ error\footnote{Force error as per Eq. (\ref{Force-Error-NT}) due to graph representation as shown in Figure \ref{fig_graph_force_comp}. Units: mEh/Bohr.} 
\end{tabular} &

\multicolumn{3}{c}{
\begin{tabular}{@{}c@{}}$\Delta F^{ML}$ error\footnote{Force error as per Eq. (\ref{Force-Error-NT}) due to machine learning approximations as shown in Figures \ref{fig_nn_force_comp} and \ref{fig_nn_force_2}. Units: mEh/Bohr.} 
\end{tabular}} \\
\cline{3-5}  
& & 10\%\footnote{Fraction of training data obtained using k-means.} & 20\%\footnotemark[3] & 40\%\footnotemark[3] \\ \hline
dynamic & 0.056 & 0.057 &  0.056 &  0.056 \\ 
1 & 0.055 & 0.060 & -  & -  \\ 
2 & 0.041 & 0.096 & -  &  - \\
3 & 0.016 & 0.132 &   0.059&  0.032 \\

\hline \hline
\end{tabular*}
\caption{Errors in fragment forces as per }
\label{tab_force_graph}
\end{table}

As the simplex rank cutoff increases, the force component distributions and MAE systematically converge. However, we also see in Figure \ref{fig_graph_force_comp}(b), that for ${\cal R}=1$, the larger magnitude forces deviate from the exact value. This is not the case for the dynamically determined edges in Figure \ref{fig_graph_force_comp}(a) and is likely due to the non-bonded edges included in Figure  \ref{fig_graph_force_comp}(b) as can be seen from Figure \ref{fig:RDF}(a). However, as we increase the value of ${\cal R}$ to 2, these are corrected (Figure \ref{fig_graph_force_comp}(c)) before eventually converging to the exact value in Figure \ref{fig_graph_force_comp}(d) for ${\cal R}=3$. 

The baseline established here, allows us to better evaluate the performance of the ML potentials in the next section. 

\begin{figure}
        \subfigure[Dynamic edge cutoff (10\% training)]{\includegraphics[width=0.48\linewidth]{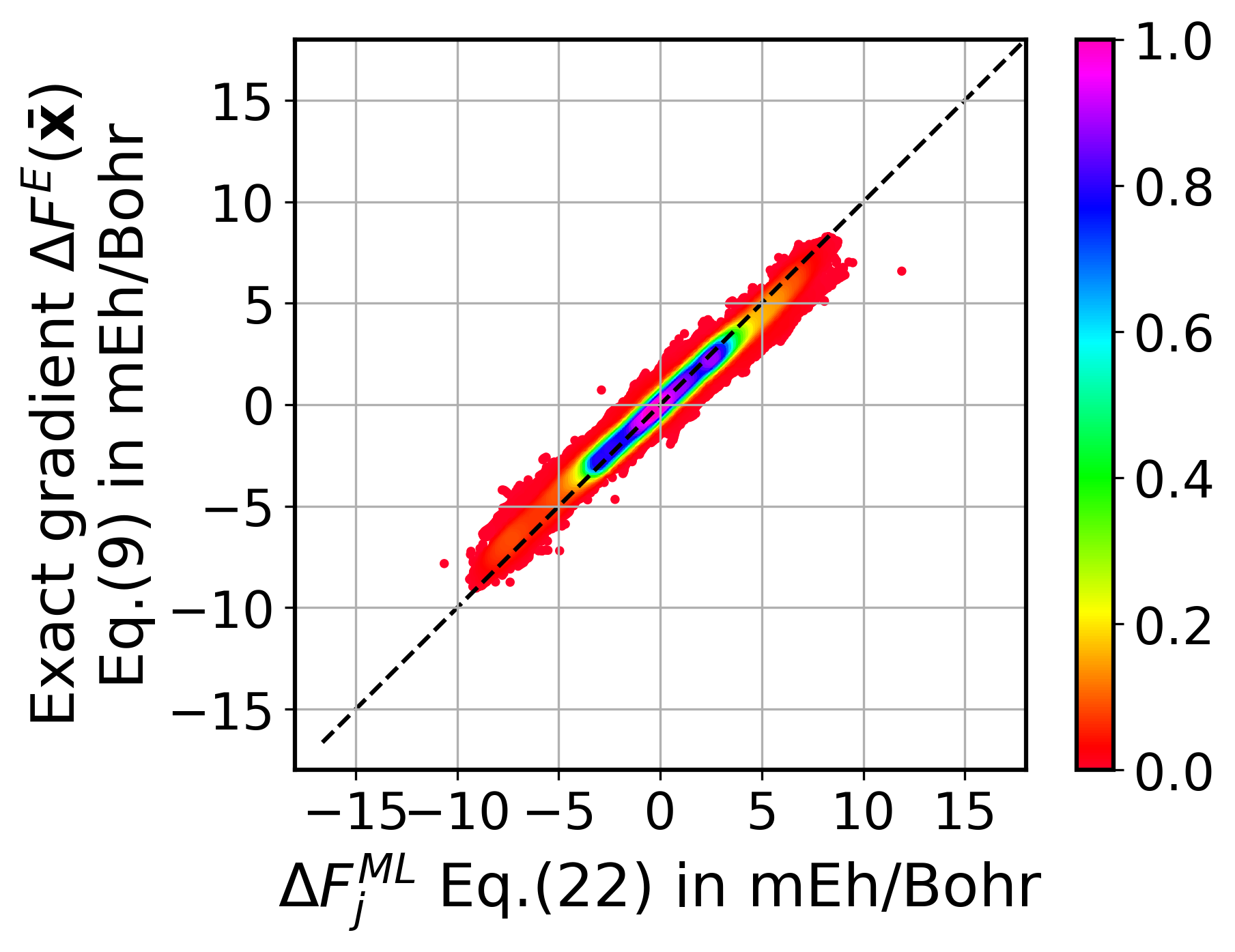}}
        \subfigure[${\cal R}=1$ (10\% training)]{\includegraphics[width=0.48\linewidth]{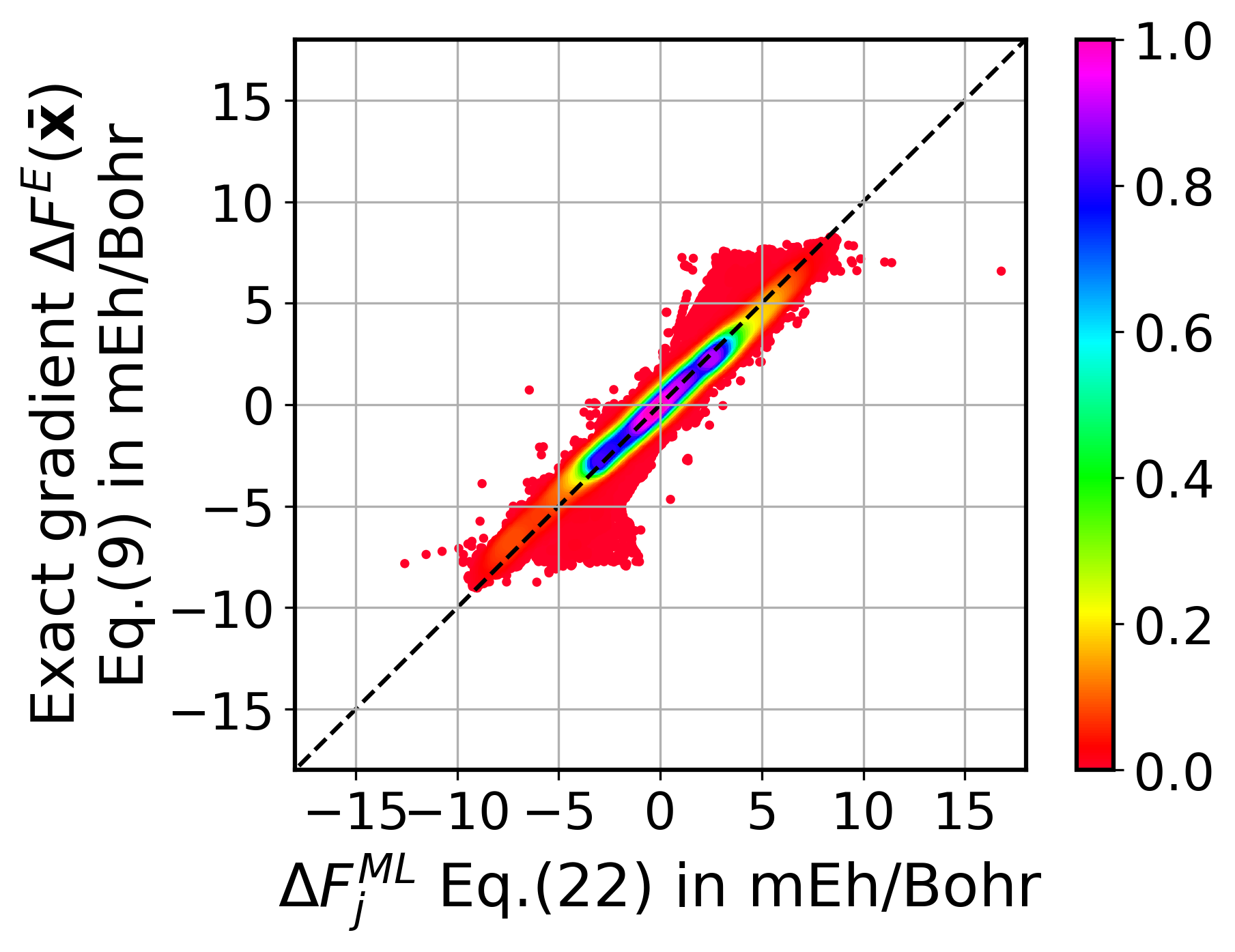}}        \subfigure[${\cal R}=2$ (10\% training)]{\includegraphics[width=0.48\linewidth]{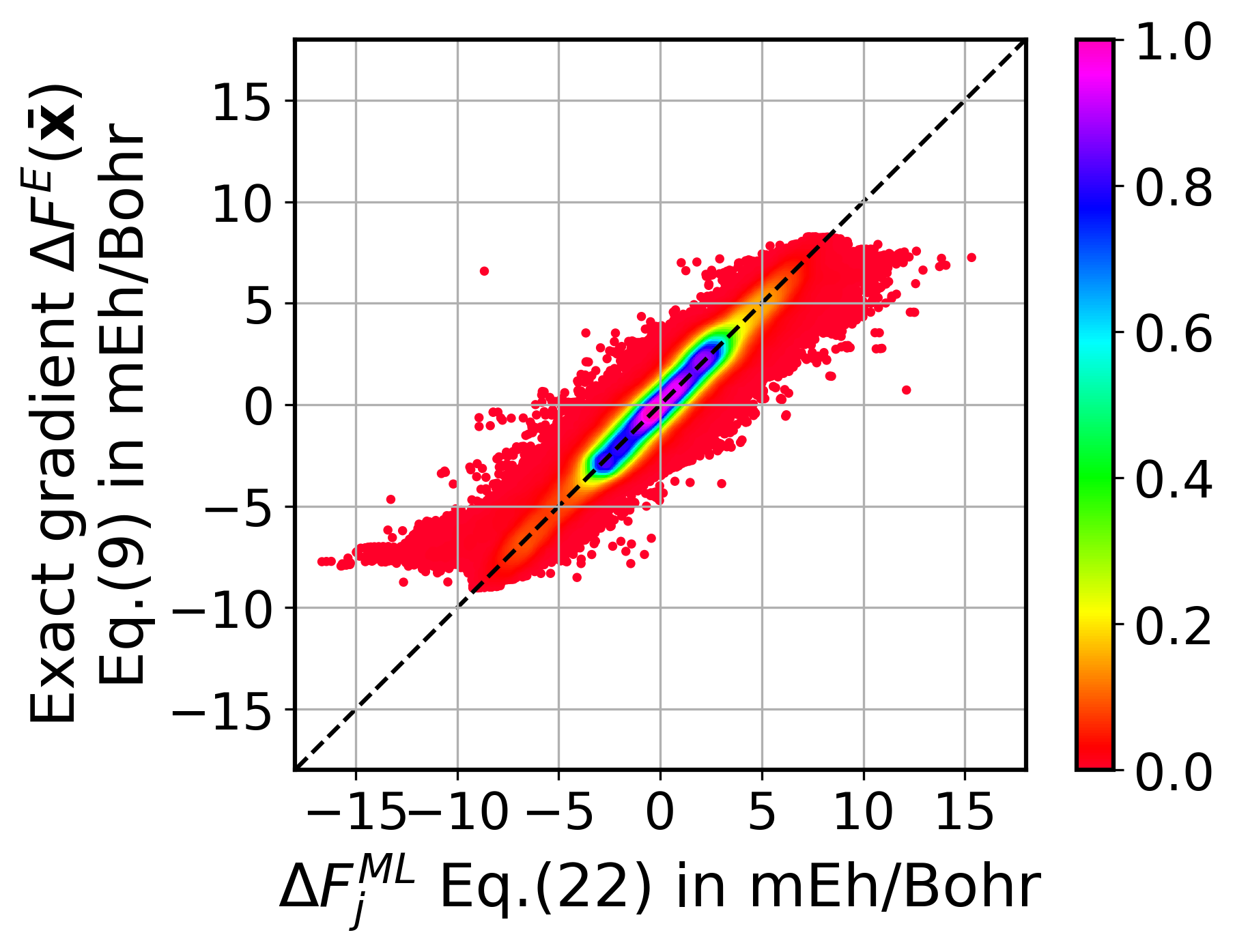}}        \subfigure[${\cal R}=3$ (10\% training)]{\includegraphics[width=0.48\linewidth]{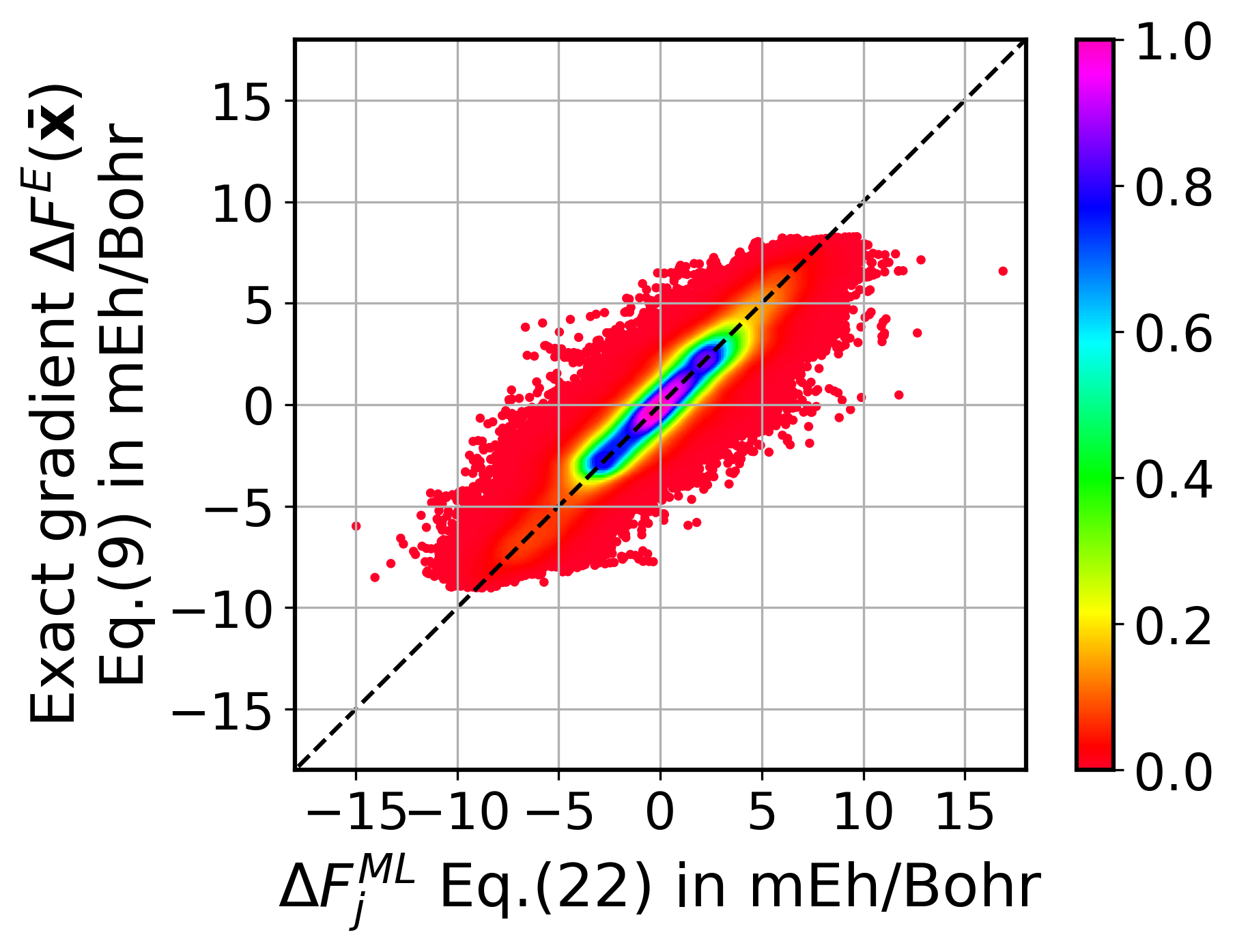}}              
    \caption{Full system force component predicted at different maximum rank cutoff for 10\% fragment training.}
    \label{fig_nn_force_comp}
\end{figure}

\begin{figure}
\subfigure[Model set 1 training with 10\% fragments at ${\cal R}=3$]{\includegraphics[width=0.48\linewidth]{figures/10_r3-2.png}} 
\subfigure[Model set 2 training with 20\% fragments at ${\cal R}=3$]{\includegraphics[width=0.48\linewidth]{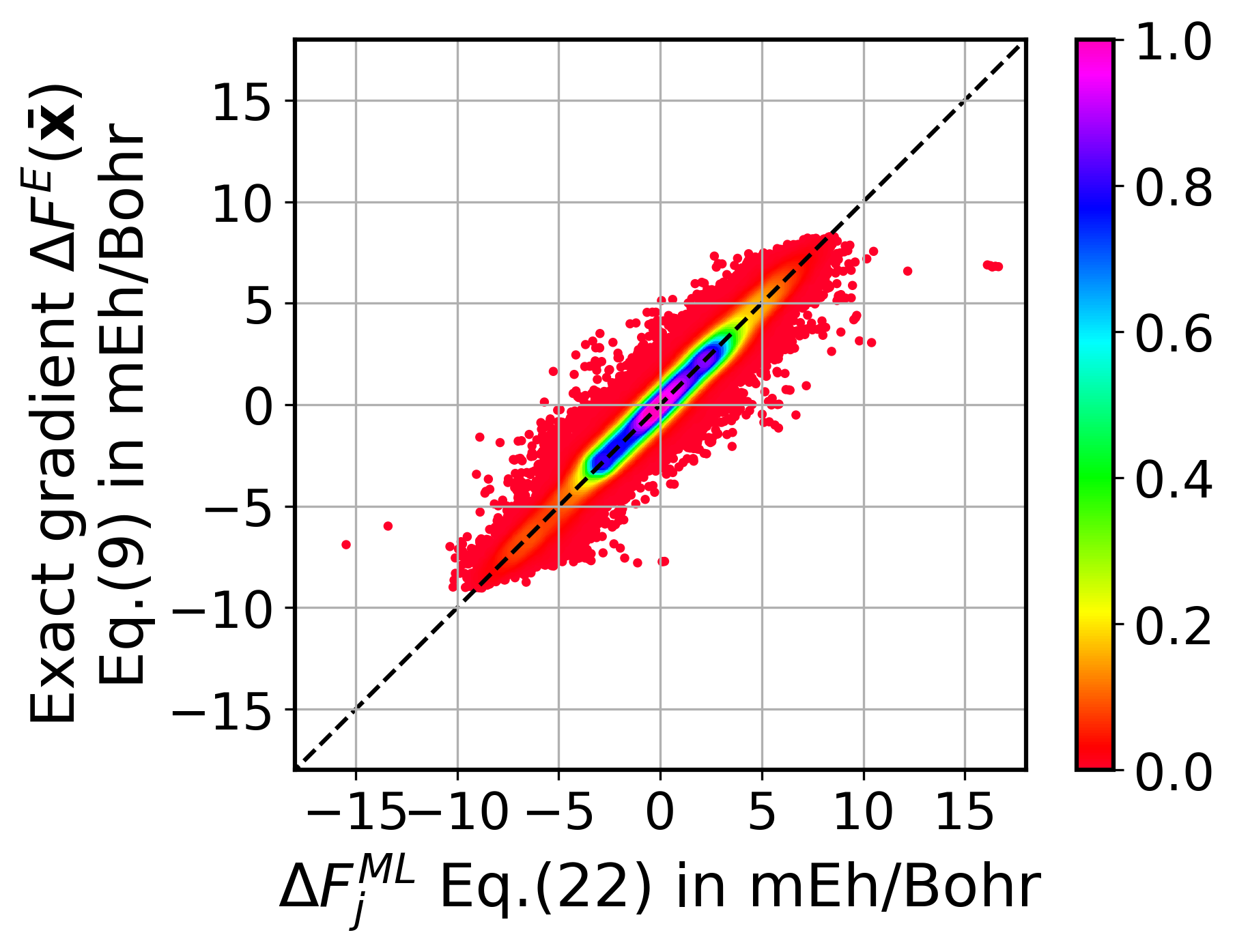}}        \subfigure[Model set 3 training with 40\% fragments at ${\cal R}=3$]{\includegraphics[width=0.48\linewidth]{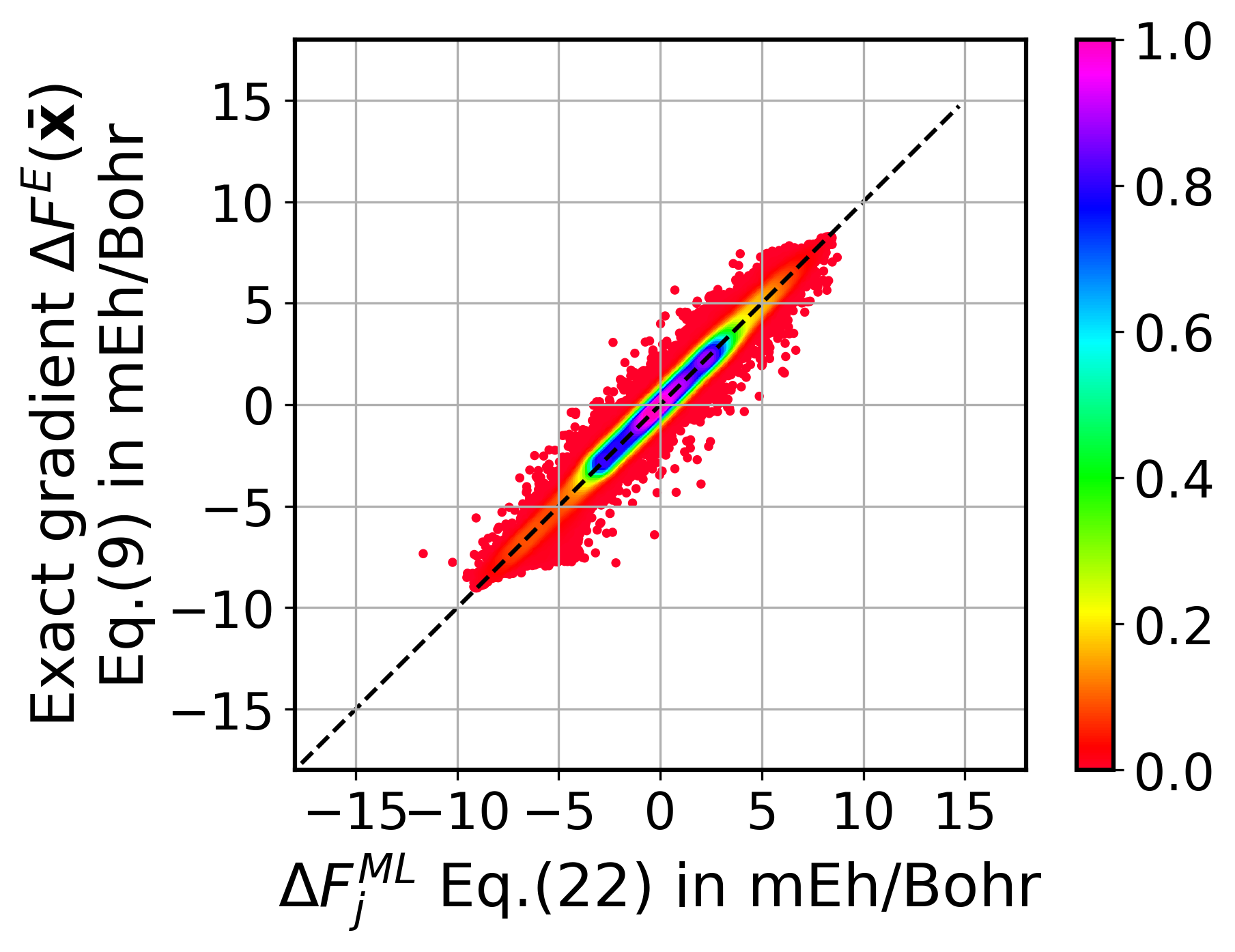}}        
    \caption{Increasing training set size systematically improve the accuracy on force components at large rank cutoff.}
    \label{fig_nn_force_2}
\end{figure}

\subsection{Full system force predictions from graph based fragment neural networks}
The error distributions for the full system forces corresponding to the sets of neural network models tested in Section \ref{sec_frag}, 
are shown in Figure \ref{fig_nn_force_comp}.
Unlike the converging trend 
in Figure \ref{fig_graph_force_comp}, the predicted component error distributions get broader for the higher rank calculations of ${\cal R}>1$ when using 10\% of the AIMD fragment datasets for training. (See Table \ref{tab_force_error} for training fractions.)
This behavior stems from the fact that 
for a fully connected graph, the population of simplexes across different ranks follows the combinatorial relationship ${6 \choose {r+1}}$ for rank $r$ simplexes. 
In the reference trajectory, the number of fragments that contain an excess proton 
depends on the position of central transferring hydrogen, and the number of fragments generated from the reference trajectory is shown Table \ref{tab_force_error}.
As we can see from  Table \ref{tab_force_error},
the number of protonated fragment species in the solvated zundel dominates for larger simplexes, resulting in the high overall population contributions from $H_7O_3^+$ and $H_9O_4^+$.
Consequently, prediction errors accumulate rapidly during the full-system reconstruction, eclipsing the physical accuracy gained by incorporating higher order terms. Additionally, as we see in Figure \ref{fig_nn_force_comp} the dynamic edge cut-off, where only hydrogen bonded rank 1 simplexes are included, shows a much more stable behavior in the accuracy of the predictions.

Since the major portion of the error originates from the neural network models for larger simplexes, increasing the training ratio serves as an effective mitigation strategy. 
Towards this, we construct models 
using 20\% and 40\% of fragments for training and the  corresponding error distributions are shown in Figures \ref{fig_nn_force_2}(b) and (c), respectively. These may be compared with Figure \ref{fig_nn_force_2}(a) where the 10\% results of \ref{fig_nn_force_comp} (d) are reproduced. Clearly, expanding the training ratio systematically lowers the component MAE by narrowing the overall variance and tightening the error distributions.

However, generating training configurations for large simplexes demands expensive, high level electronic structure calculations, while also demanding more complex network architectures with extended training epochs. In the context of on-the-fly training and active learning applications, these computational bottlenecks severely affect the applicability 
of machine learning-accelerated molecular dynamics. The dynamic edge distance provides an elegant compromise between accuracy and efficiency. While prior publications like Ref \cite{frag-ML-Xiao,Xiao-LLM} revealed that potential energy surfaces truncated at small static rank cutoffs suffer from systematic energy shifts, the force prediction results presented here do not suffer from these deficiencies. The dynamic rank 1 framework provides a tight and rigorous alignment of error distributions, delivering good accuracy without introducing systematic biases toward either overestimation or underestimation. 

\begin{figure}
\subfigure[oxygen-oxygen distance distribution]{\includegraphics[width=0.49\linewidth]{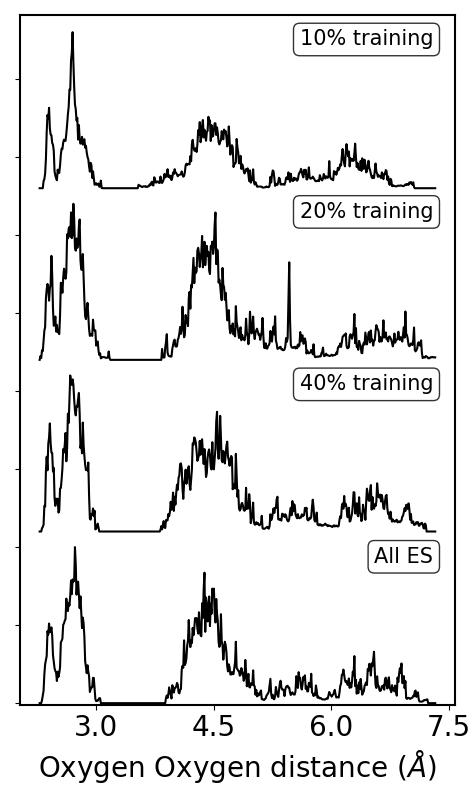}}
\subfigure[oxygen-oxygen-oxygen angle distraibution]{\includegraphics[width=0.49\linewidth]{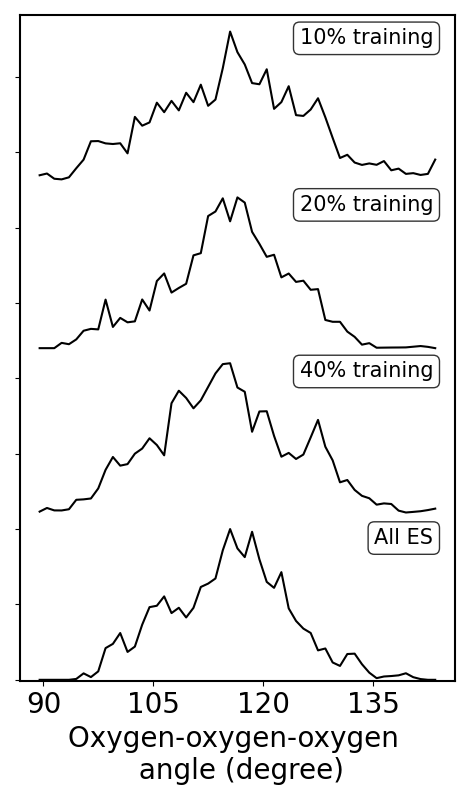}}
\caption{Oxygen-oxygen distance and oxygen-oxygen-oxygen angle distribution comparison between the reference trajectory (top) and a fully predicted trajectory (bottom).}
    \label{fig_oo}
\end{figure}

\begin{figure}
\subfigure[Oxygen-hydrogen distance distribution]{\includegraphics[width=0.49\linewidth]{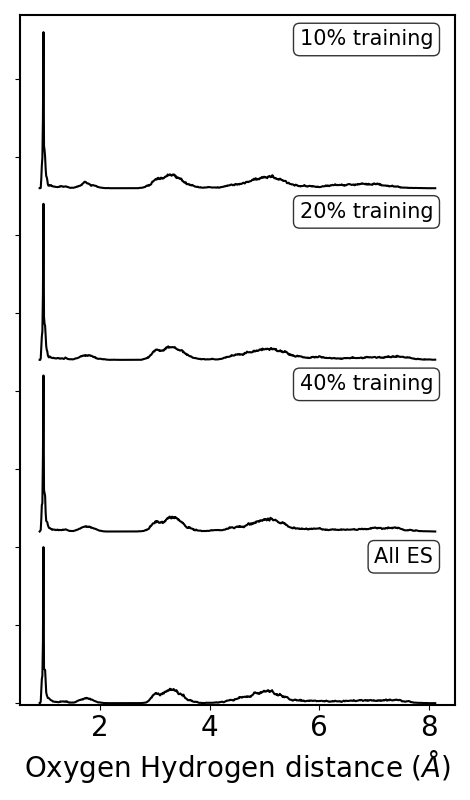}}
\subfigure[Oxygen-hydrogen distance distribution $\geq 1.4\AA$]{\includegraphics[width=0.49\linewidth]{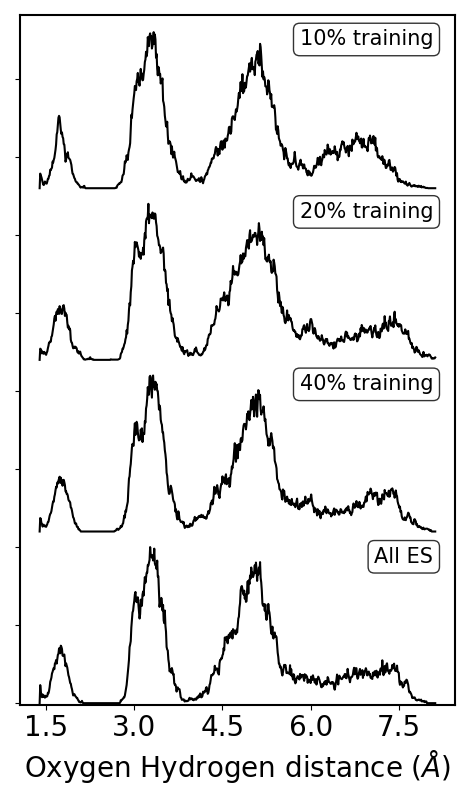}}
\caption{Oxygen-hydrogen distance distribution comparison between the reference trajectory and a fully predicted trajectories from models trained with different training set size).}
    \label{fig_oh}
\end{figure}

\begin{figure}
\subfigure[Velocity correlation]{\includegraphics[width=0.49\linewidth]{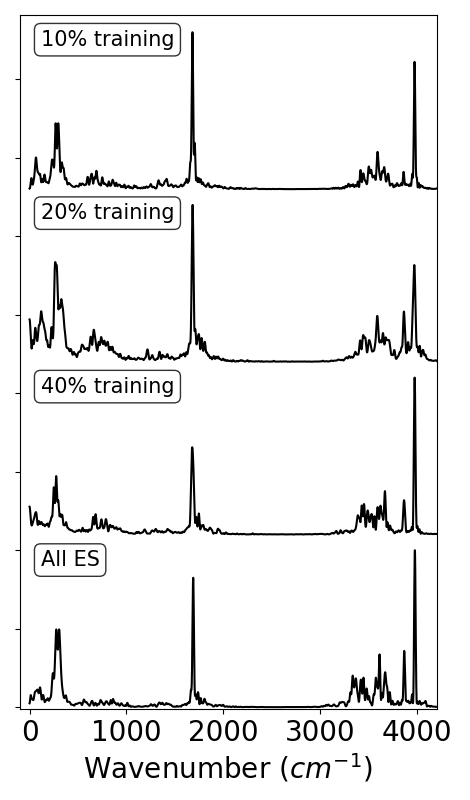}}
\subfigure[OH stretch region]{\includegraphics[width=0.49\linewidth]{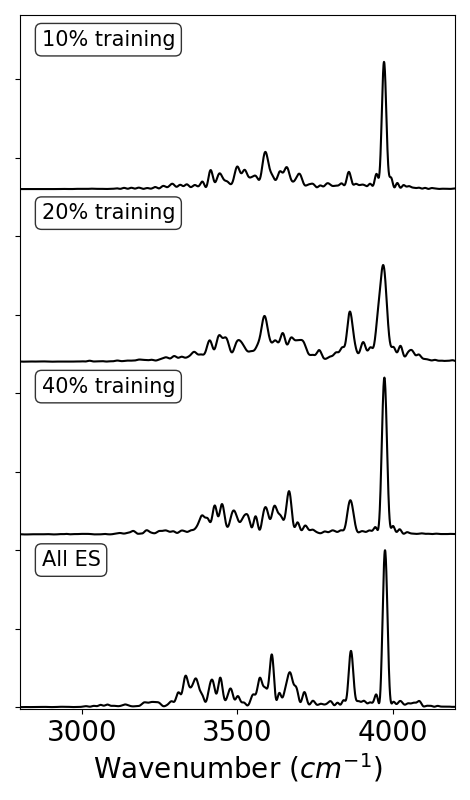}}
\caption{Velocity correlation comparison between the reference trajectory and a fully predicted trajectories from models trained with different training set size.}
    \label{fig_ftvac}
\end{figure}



\section{Accuracy of AIMD trajectories from machine-learning  based CCSD trajectories}\label{sec_autocor}
To evaluate the practical performance of the dynamic distance edge cutoff with ML, we generated fully predicted trajectories using the force models trained on the 10\%, 20\% and 40\% fragments 
as introduced earlier in Figures \ref{fig_nn_force_comp} and \ref{fig_nn_force_2}. These trajectories was initialized with the same geometry and kinetic energy as a reference trajectory for CCSD obtained using forces in Eq. (\ref{eq_graph-f}). The structural features obtained from dynamics are discussed in Section \ref{RDF-sec} and the dynamical features in Section \ref{FTVAC-sec}.

\subsection{Structural features}
\label{RDF-sec}
The resulting oxygen-oxygen and oxygen-hydrogen structural distributions are provided in Figure \ref{fig_oo} and \ref{fig_oh}. As shown in the figures, the peak positions align extremely well between the trajectories, demonstrating that the primary structures are accurately captured in the machine-learning-based dynamics trajectories. 

However the predicted trajectory using 10\% fragments shows a tendency of shrinking the second peak at 2.7{\AA} and yields a broader distribution around 4{\AA} as compared to the reference trajectory labeled as ``All ES'' at the bottom. This artificial broadening stems from the fact that the purely data driven force models lack explicit physical constrains. Consequently, as localized force error sequentially accumulates over time, the trajectory gradually shifts to sparsely sampled or previously unvisited regions of the potential energy surface, degrading the quality of trajectory over time. This problem is mitigated when we increase the training set size. With 40\% fragments trained, the predicted trajectory aligns better with the reference trajectory for both peak positions and shape. However, this also implies that 
an incremental transformation of the neural network models, as described in Ref. \cite{Xiao-LLM} may be necessary for forces and this will be considered in future publications. 

\subsection{Dynamical features}
\label{FTVAC-sec}
We also compute the Fourier transform of the velocity auto-correlation function as per
\begin{align}
I_V(\omega) =& \lim_{T\rightarrow\infty} \int_{t=0}^{t=T} dt e^{-\imath \omega t} \langle {\bf V(0)} \cdot {\bf V(t)} \rangle
\nonumber \\ =&
 \lim_{T\rightarrow\infty} \sum_{i=1}^{N_{Atoms}} \sum_{j=1}^3 \int_{t=0}^{t=T} dt e^{-\imath \omega t} \nonumber \\ & \quad \quad \quad \int_{t^\prime=0}^{t^\prime=T} dt^\prime
    V_{i,j}(t^\prime) * V_{i,j}(t^\prime+t) \nonumber \\ =&
\lim_{T\rightarrow\infty} \sum_{i=1}^{N_{Atoms}} \sum_{j=1}^3 {\left|
  \int_{t=0}^{t=T} dt e^{-\imath \omega t} V_{i,j}(t) \right|}^2,
\label{FTVACdefn}
\end{align}
where the term $\langle \cdots \rangle$, in the first equation, indicates ensemble average and is equal
to the $t^\prime$-integral in the second equation under the ergodicity condition. We have used the
convolution theorem \cite{Numerical-Recipes} to
reduce the second equation to the third equation. The quantity, $I_V(\omega)$ provides the
vibrational density of states and is shown in Figure \ref{fig_ftvac}. 

\begin{figure}
    {\includegraphics[width=0.9\linewidth]{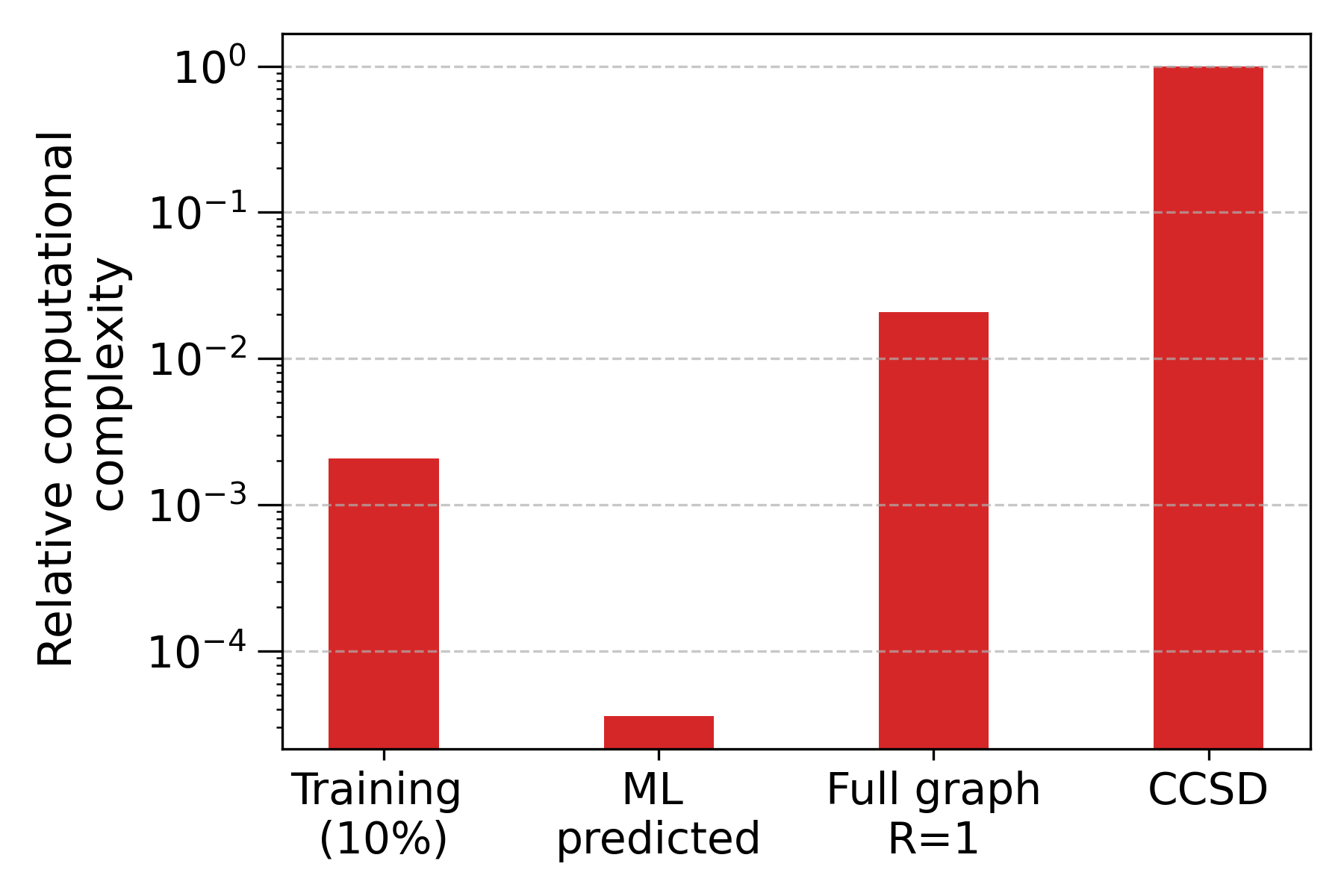}}
    \caption{Relative computational complexity for training sets, ML corrected trajectories, reference trajectory and CCSD trajectories.}
    \label{fig_complexity}
\end{figure}

Clearly, the machine learning based AIMD trajectories correctly obtain all features in the full trajectory. The corresponding CPU time investment needed to generate such trajectories are shown in Figure \ref{fig_complexity}. Here we 
include training costs (left vertical bar) since these are integral to the learning model. However, the key idea is that the graph protocol allows one to treat the training problem as being independent of system size. In fact, the training effort is system size independent as seen in Ref. \cite{Xiao-LLM} and hence the relative scale ratios between the two vertical bars on the right can be much larger for larger systems. The ML calculations are and extremely small fraction of the cost while preserving accuracy. Clearly while some care is needed in generating accurate models, the promise of generating AI-based MD trajectories is very much a reality. 



\section{Conclusion}\label{sec_conclusion}
We have presented a robust, graph-theoretic molecular fragmentation framework integrated with machine learning to directly model post-Hartree-Fock nuclear forces and enable highly accurate AIMD simulations at coupled cluster accuracy. By addressing the severe computational scaling of correlated electronic structure methods, this framework offers a highly scalable and general solution to the simulation of complex, fluxional chemical systems.  

The cornerstone methodology of this work is built upon several key pillars: (a) Direct Force Vector Learning: Rather than relying on automatic differentiation of a learned energy surface, which struggles in sparsely sampled regions and is severely confounded by link-atom Jacobians, in fragmented models—our neural networks directly predict nuclear force vectors. (b) Covariant Principal-Axis Descriptors: By projecting atomic force vectors onto the fragment-fixed principal axes of inertia, we establish directionally aware, standardized target blocks that naturally preserve rotational, translational, and permutational invariance. (c) High Parameter Efficiency: Our vector-valued training protocol shares abstract basis functions across force components, recovering the accuracy of scalar component-wise models while reducing the total trainable parameter count by over an order of magnitude. (d) Unsupervised Space Tessellation: Utilizing a mini-batch $k$-means clustering algorithm to partition the multidimensional geometry space of molecular fragments, we balance structural variations across all degrees of freedom to construct representative training databases with only 10\% to 20\% of the reference configurations. 

Validation and dynamical accuracy studies for the framework was carried out on the highly fluxional solvated Zundel cation ($H_{13}O_{6}^+$), which presents a critical test due to its strong polarization, charge delocalization, and complex proton transfer dynamics. Furthermore, this moiety is central to the study of proton transfer in atmospheric, biological, materials and condensed phase systems. While static, fully connected graphs show error accumulation at higher simplex ranks under limited training ratios, our dynamic edge cutoff (rank-1) framework delivers a highly stable and computationally elegant compromise. Fully machine-learning-predicted AIMD trajectories successfully reproduced key structural characteristics such as the radial distribution functions for $\text{O–O}$ and $\text{O–H}$, and complex dynamical signatures, including the velocity autocorrelation power spectrum (vibrational density of states).  

Looking ahead, this approach effectively bridges the gap between high-level correlated wavefunction theories and the long timescales required to sample reactive chemical systems. Because our graph-theoretic reconstruction shares a deep mathematical connection with attention-based Large Language Model (LLM) architectures, the direct force modeling scheme is naturally poised to leverage advanced transfer-learning protocols. This will enable pre-trained localized fragment models to scale dynamically to larger, highly polarized, and reactive condensed-phase environments, laying down a highly efficient, general, and systematically improvable path for modern chemical dynamics simulations.

\section{Acknowledgment}
This research was supported by the National Science
Foundation grant CHE-2102610 (along with the associated Creativity Extension) to SSI. The computational facilities at Indiana University are duly acknowledged and have been critical to this effort. The acknowledgment is due in part to the Lilly Endowment, Inc., for their support of the BigRed computing facility at Indiana University widely used in the effort represented in this publication. This work was also supported in part by Shared University Research grants from IBM, Inc., to Indiana University, which supports the Scholarly Data Archives. 


\appendix

\section{Connections between the graph-theoretic ML models and Large language models facilitated transfer learning in Ref \cite{Xiao-LLM}}
\label{LLM}
In Ref. \cite{Xiao-LLM}, we discuss the mathematical connections between the graph theoretic formalism presented above and the concept of attention-based aggregation in large language model (LLM) architectures\cite{trans-attention}. These mathematical connections provide a recipe for transfer-learning protocols\cite{Xiao-LLM}, and form the conceptual basis for the formalism here for nuclear dynamics. In essence, in modern natural language processing (NLP) words and phrases are embedded as vectors that encapsulate relationships among neighboring linguistic components based on semantics. A similar abstraction arises in our graph-theoretic representation of molecular potential energy surfaces where fragment-fragment interactions are embedded in simplex energies. In both cases, the role of an individual component is determined by its environment: the context of neighboring words in language and the surrounding graph structure of molecular fragments here. For example, the graph-theoretic coefficients ${\cal M}^{\cal R}_{\alpha_r,r}$ in Eq. (\ref{eq_deltaE}) determine the importance of each fragment and similarly, in linguistics, the so-called {\em softmax} function variable defined as 
\begin{align}
\left[ softmax\left(\frac{Q K^T}{\sqrt{d}}\right)\right]_{i,j} = \frac{\exp{{Q_i\cdot K_j}/\sqrt{d}}}{\sum_{k}\exp{{Q_i\cdot K_k}/\sqrt{d}}}
\label{softmax-eq}
\end{align}
assigns importance to each {\em key-value} pair (vectors $K$ above for {\em key} and $V$ below for {\em value})  based on its relevance to a {\em query} (vector $Q$ above). These weights in NLP play a role analogous to the graph-theoretic coefficients in our graph-theoretic molecular fragmentation formalism and define the critical term ``attention'' as
\begin{align}
    Attention(Q,K,V) = softmax\left(\frac{Q K^T}{\sqrt{d}}\right)V
    \label{eq_attention}
\end{align}
which turns out to have analogous meaning as $\Delta E({\bf {\bar x}})$  in Eq (\ref{eq_deltaE}). Thus, 
\begin{align}
{\text {Importance}} \equiv \left[ softmax\left(\frac{Q K^T}{\sqrt{d}}\right)\right]_{i,j} \leftrightarrow {\cal M}_{\alpha_r, r}^{\cal R}
\label{analogy-1}
\end{align}
Likewise, the transformer value vectors correspond naturally to the fragment energy corrections,
\begin{align}
{\text {Intrinsic Value}} \equiv V_{j,l} \leftrightarrow \Delta E_{\alpha_r,r}({\bf {\bar x}})
\label{analogy-2}
\end{align}
and finally
\begin{align}
{\text {Attention (Q,K,V) }} \leftrightarrow  \Delta E({\bf {\bar x}})
\label{analogy-3}
\end{align}
Viewed in this way, both frameworks perform an importance-weighted aggregation of local information to construct a global description. In NLP, this aggregation yields contextual meaning; in molecular systems, it yields the stability and energy of a molecular configuration. 

Despite these similarities, the primary challenge differs between the two fields. In language models, attention weights must be learned from large text corpora. In molecular systems, the graph weights are determined directly from geometry and topology, while the principal challenge lies in obtaining accurate fragment energies from expensive electronic-structure calculations. Nevertheless, both approaches rely on decomposing a complex global problem into interacting local components whose weighted combination determines the final result.

In Ref. \onlinecite{Xiao-LLM} we exploited this connection to build a transfer learning protocol of molecular potential energy surfaces. To achieve this, we show that a family of neural networks used to train the quantities, $\left\{ \Delta E_{\alpha_r,r}^{ML, 1}({\bf {\bar x}_{\alpha_r,r}}) \right\}$ and obtain an accurate potential energy surface for one system, that is $\left\{ \Delta E_{\alpha_r,r}^{ML, 1}({\bf {\bar x}_{\alpha_r,r}}) \right\} \rightarrow E_{{\text{system,1}}}^{target}({\bf {\bar x}})$, 
can be refined in a well-defined and consistent numerical manner to yield an accurate potential for a much larger system, that is, 
\begin{align}
\Delta E_{\alpha_r,r}^{ML, 1}({\bf {\bar x}_{\alpha_r,r}}) \xrightarrow{\text{refine}}  \Delta E_{\alpha_r,r}^{ML, 2}({\bf {\bar x}_{\alpha_r,r}})
\label{refine}
\end{align}
and $\left\{ \Delta E_{\alpha_r,r}^{ML, 2}({\bf {\bar x}_{\alpha_r,r}}) \right\} \rightarrow E_{{\text{system,2}}}^{target}({\bf {\bar x}})$. 
This framework naturally enables transfer learning along selected subspaces of the full coordinate space, ($\mathbf{\bar{x}}$). In AIMD simulations, the decomposition of the molecular coordinates into fragment coordinates, $(\mathbf{\bar{x}}_{\alpha_r,r})$, allows geometric changes within different regions of the configurational space to be monitored separately. Since each fragment is represented by its own neural network, refinement of the potential energy surface can be focused on only those fragments that explore previously unseen regions of configuration space, as one moves from a smaller system to a larger system, and thereby avoid expensive retraining at the level of the full system.

\section{Force model accuracy for a variety of geometries}\label{sec_division}

\begin{figure}
\subfigure[H$_4$O$_2$ all fragments]{\includegraphics[width=0.23\textwidth]{figures/rank1_rank2_model_accuracy_vs_oxygen-oxygen_distance/H4O2_all3.png}}
\subfigure[H$_4$O$_2$ with O-O distance $\leq$ 3.5\AA]{\includegraphics[width=0.23\textwidth]{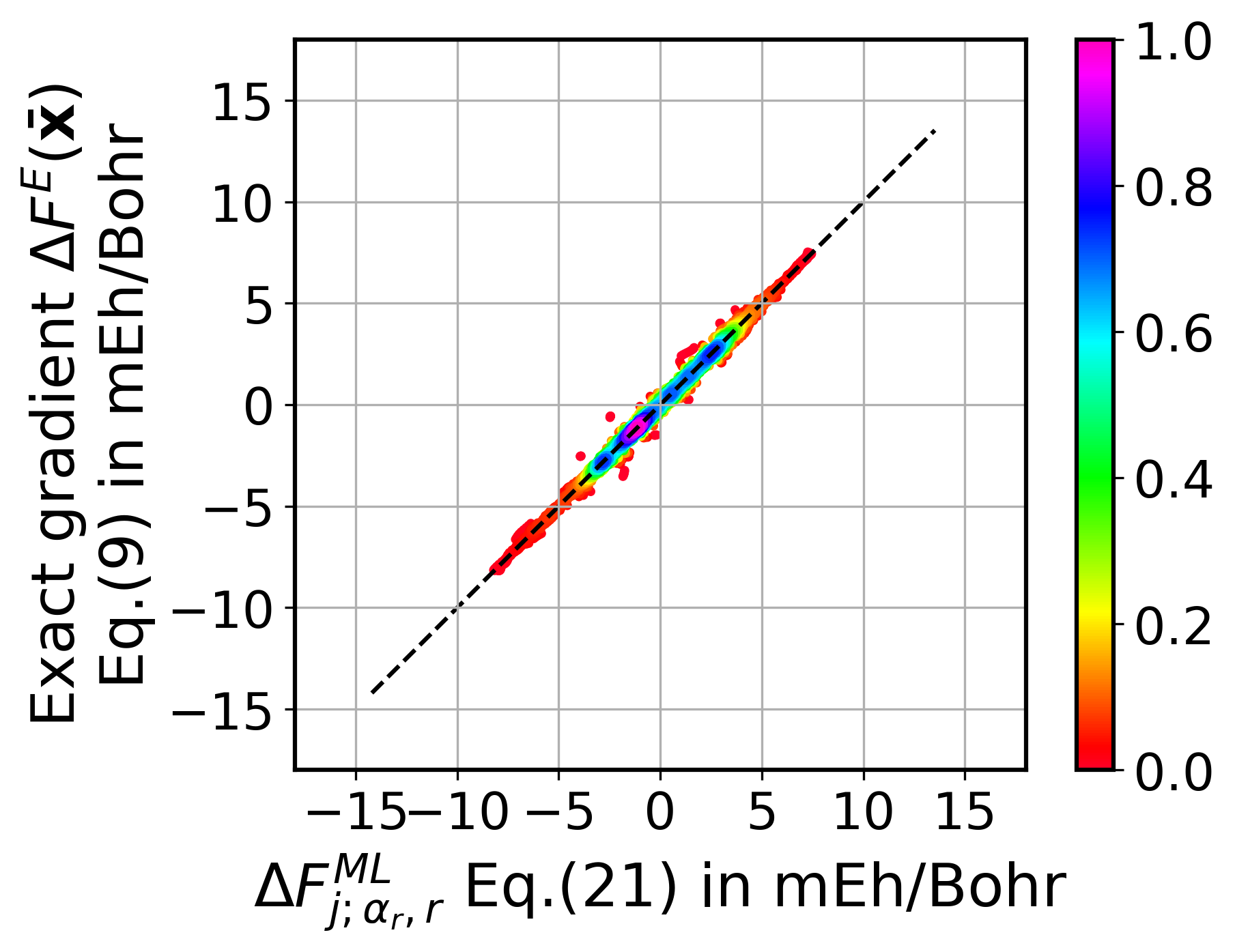}}
\subfigure[H$_4$O$_2$ with O-O distance between 3.5 and 4.5\AA]{\includegraphics[width=0.23\textwidth]{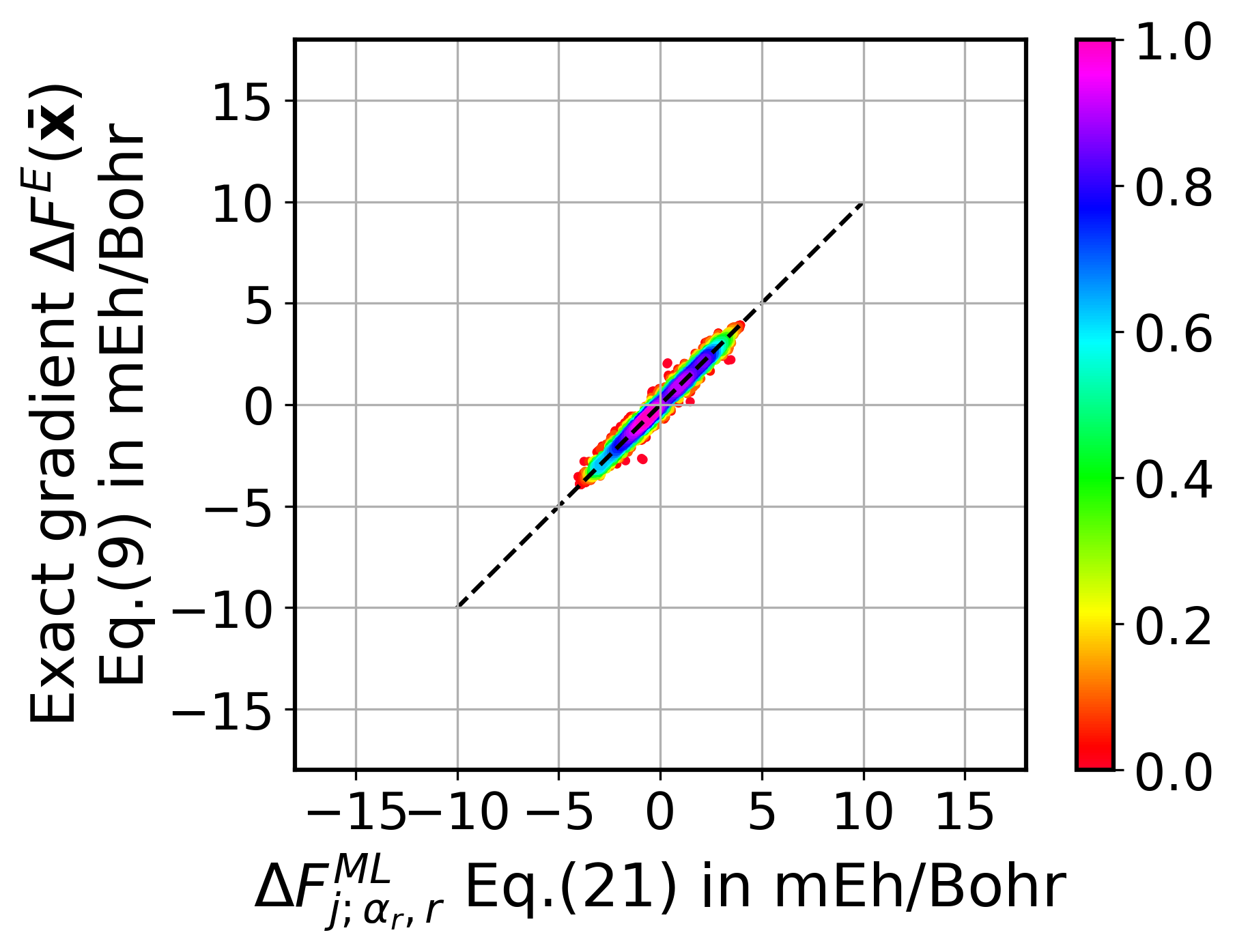}}
\subfigure[H$_4$O$_2$ with O-O distance $\geq$ 4.5\AA]{\includegraphics[width=0.23\textwidth]{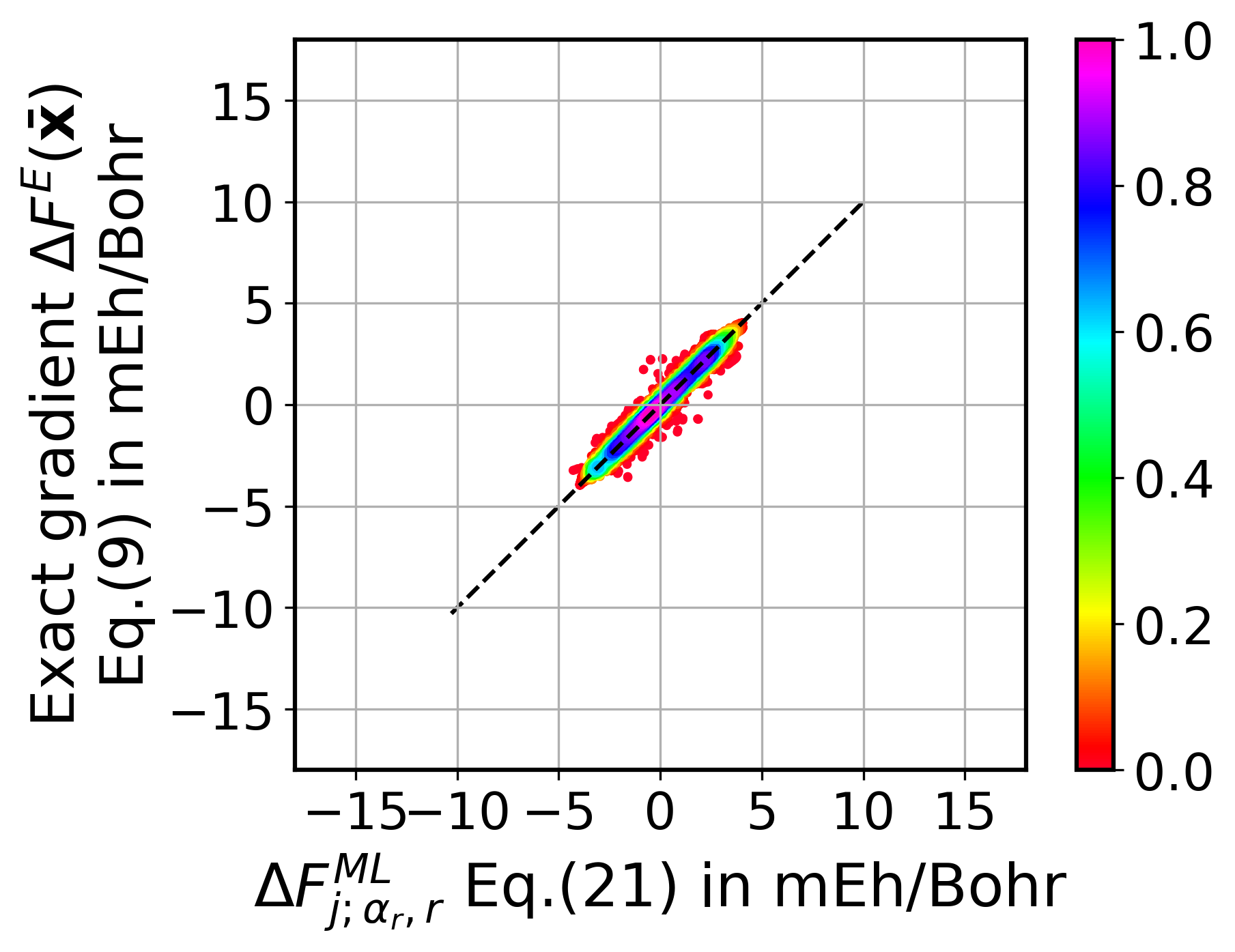}}

\caption{\label{fig_h4o2err} Distribution of fragment force errors for the given the range of oxygen-oxygen distance for $H_4O_2$.}
\end{figure}

\begin{figure}
\subfigure[H$_5$O$_2^+$ all fragments]{\includegraphics[width=0.23\textwidth]{figures/rank1_rank2_model_accuracy_vs_oxygen-oxygen_distance/H5O2_all3.png}}
\subfigure[H$_5$O$_2^+$ with O-O distance $\leq$ 3.5\AA]{\includegraphics[width=0.23\textwidth]{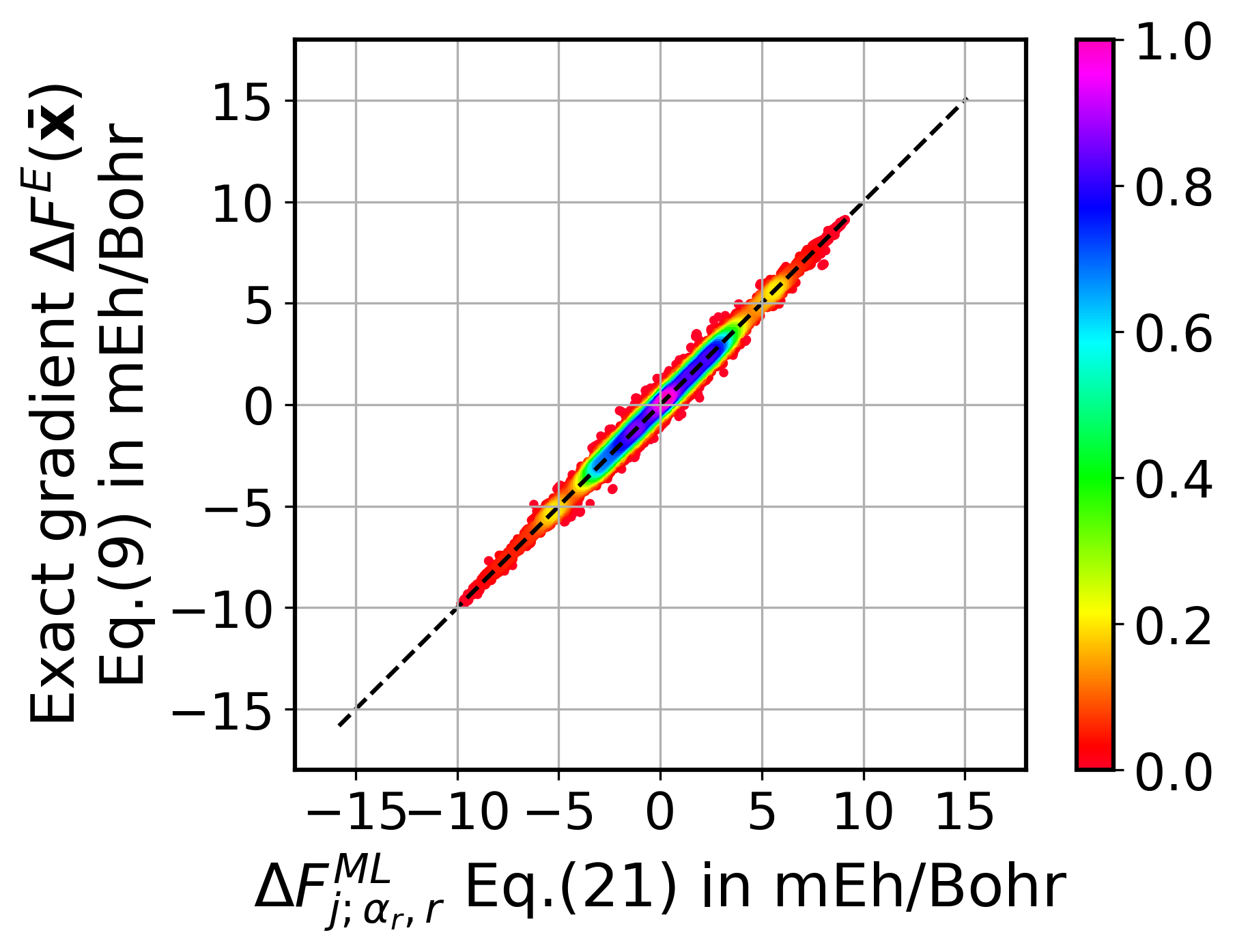}}
\subfigure[H$_5$O$_2^+$ with O-O distance between 3.5 and 4.5\AA]{\includegraphics[width=0.23\textwidth]{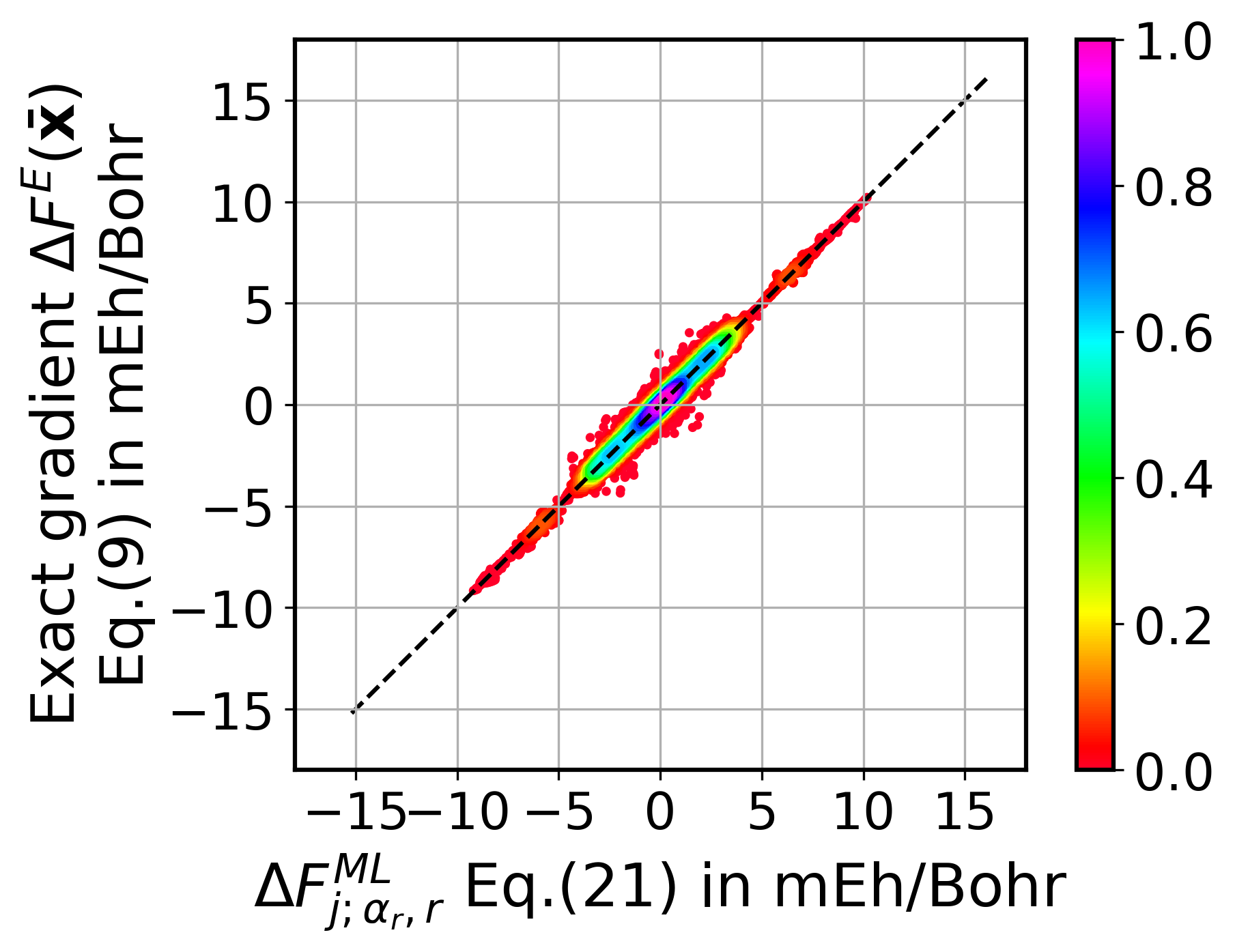}}
\subfigure[H$_5$O$_2^+$ with O-O distance $\geq$ 4.5\AA]{\includegraphics[width=0.23\textwidth]{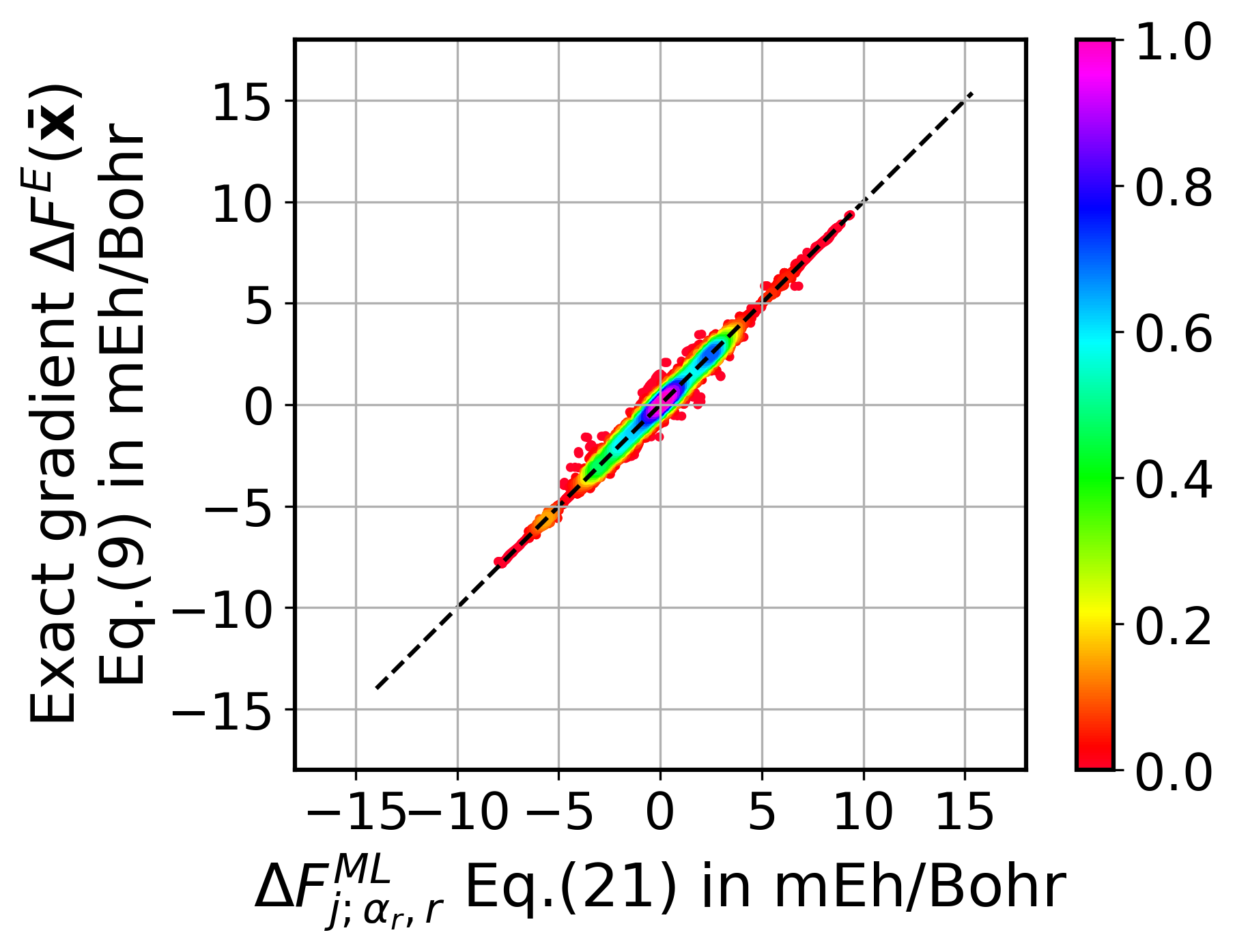}}

\caption{\label{fig_h5o2err} Distribution of fragment force errors for the given the range of oxygen-oxygen distance for $H_5O_2^+$.}
\end{figure}

\begin{figure}
\subfigure[H$_6$O$_3$ all fragments]{\includegraphics[width=0.23\textwidth]{figures/rank1_rank2_model_accuracy_vs_oxygen-oxygen_distance/H6O3_all3.png}}
\subfigure[H$_6$O$_3$ with O-O-O angles $\leq$ 90$^\circ$]{\includegraphics[width=0.23\textwidth]{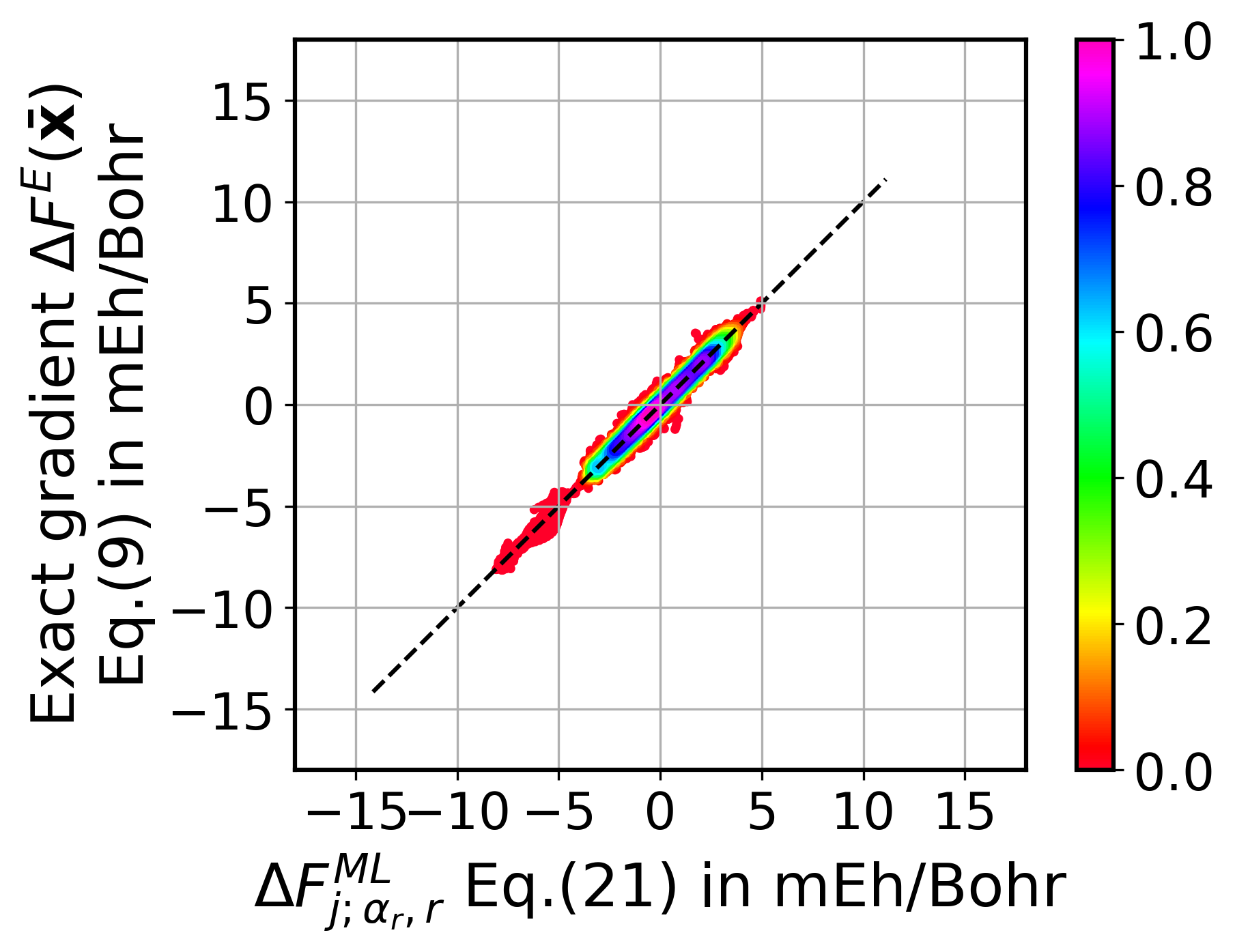}}
\subfigure[H$_6$O$_3$ with O-O-O angles between 90 and 145$^\circ$]{\includegraphics[width=0.23\textwidth]{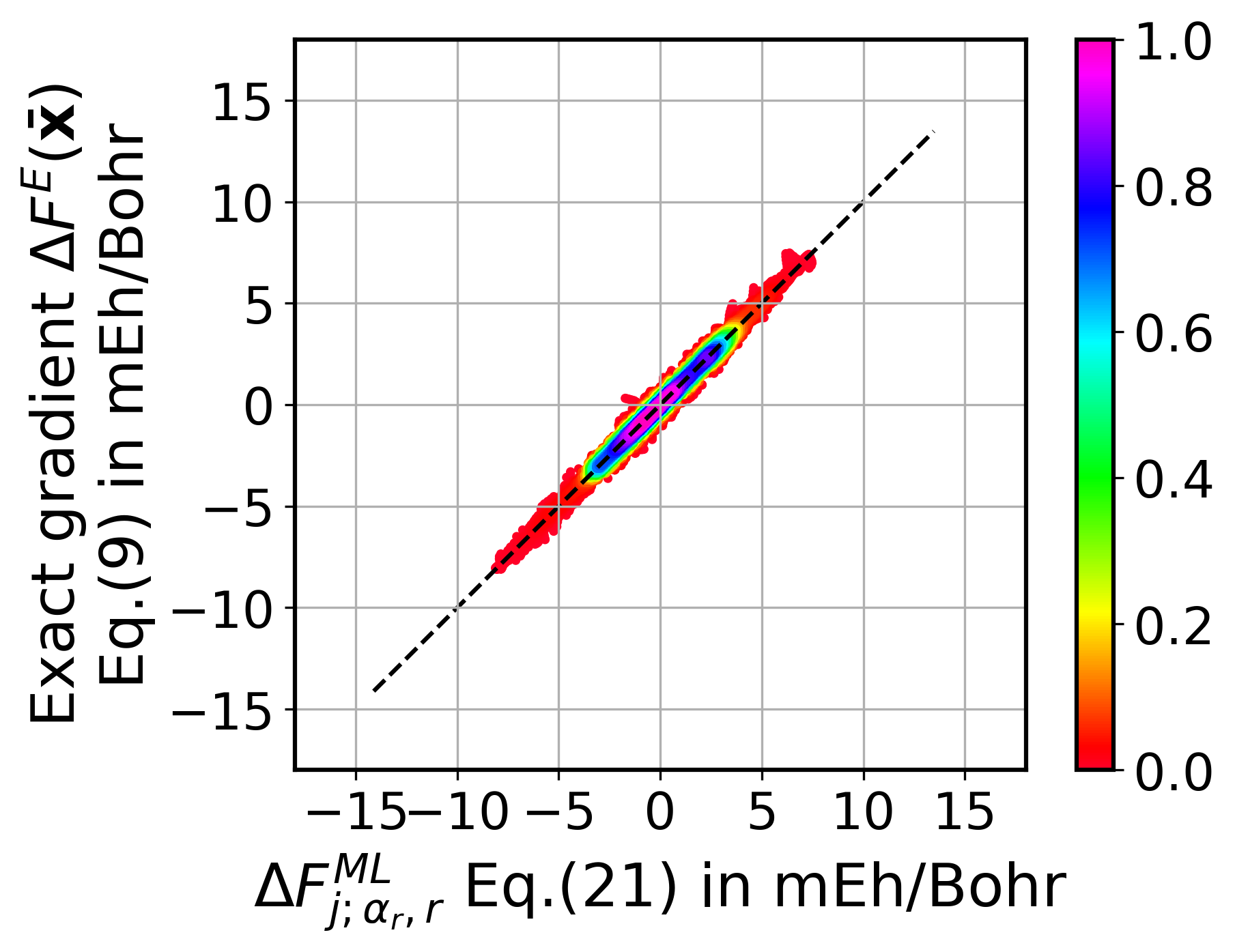}}
\subfigure[H$_6$O$_3$ with O-O-O angles $\geq$ 145$^\circ$]{\includegraphics[width=0.23\textwidth]{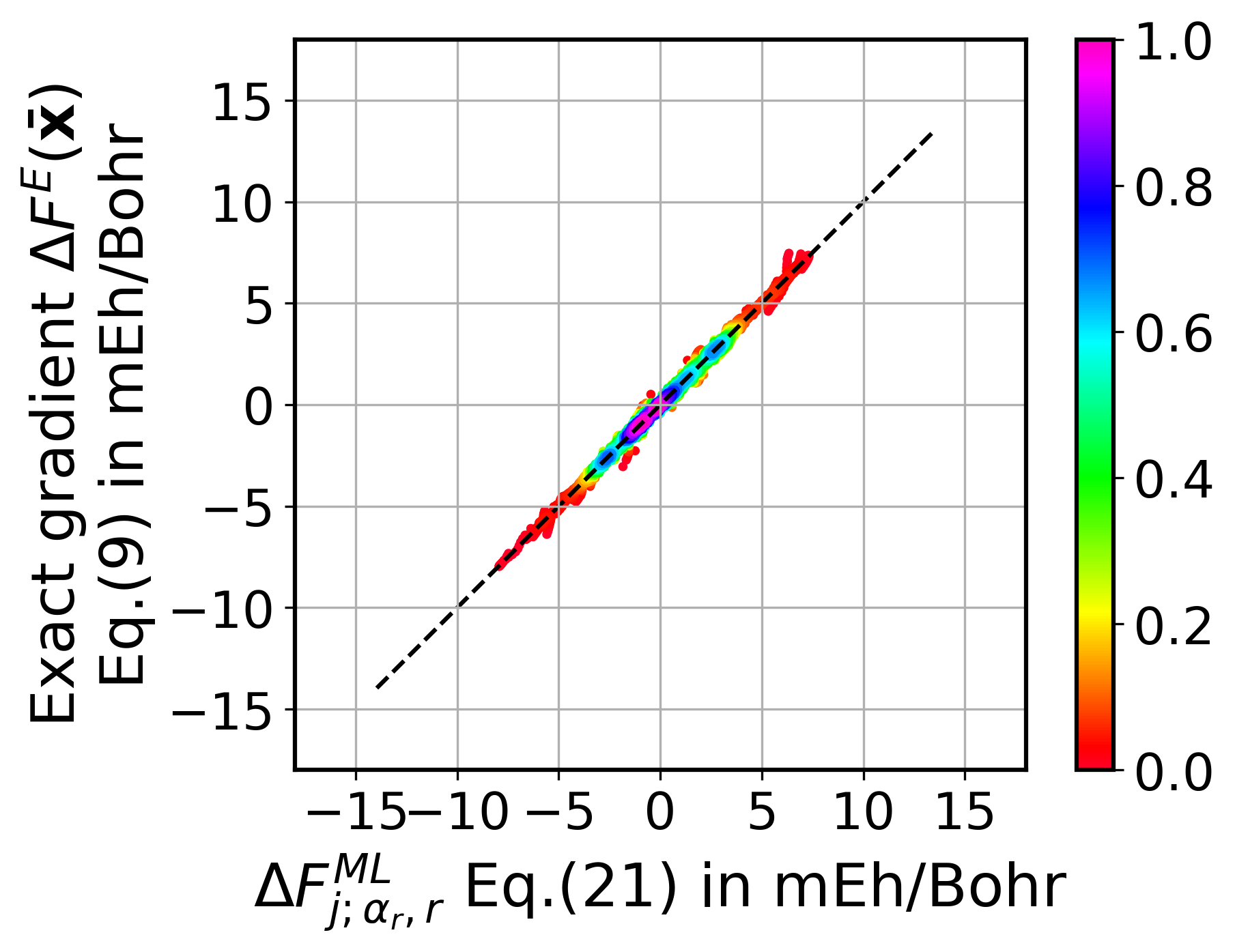}}

\caption{\label{fig_h6o3err} Distribution of fragment force errors for the given the range of oxygen-oxygen distance for $H_6O_3$.}
\end{figure}

\begin{figure}
\subfigure[H$_7$O$_3^+$ all fragments]{\includegraphics[width=0.23\textwidth]{figures/rank1_rank2_model_accuracy_vs_oxygen-oxygen_distance/H7O3_all3.png}}
\subfigure[H$_7$O$_3^+$ with O-O-O angles $\leq$ 90$^\circ$]{\includegraphics[width=0.23\textwidth]{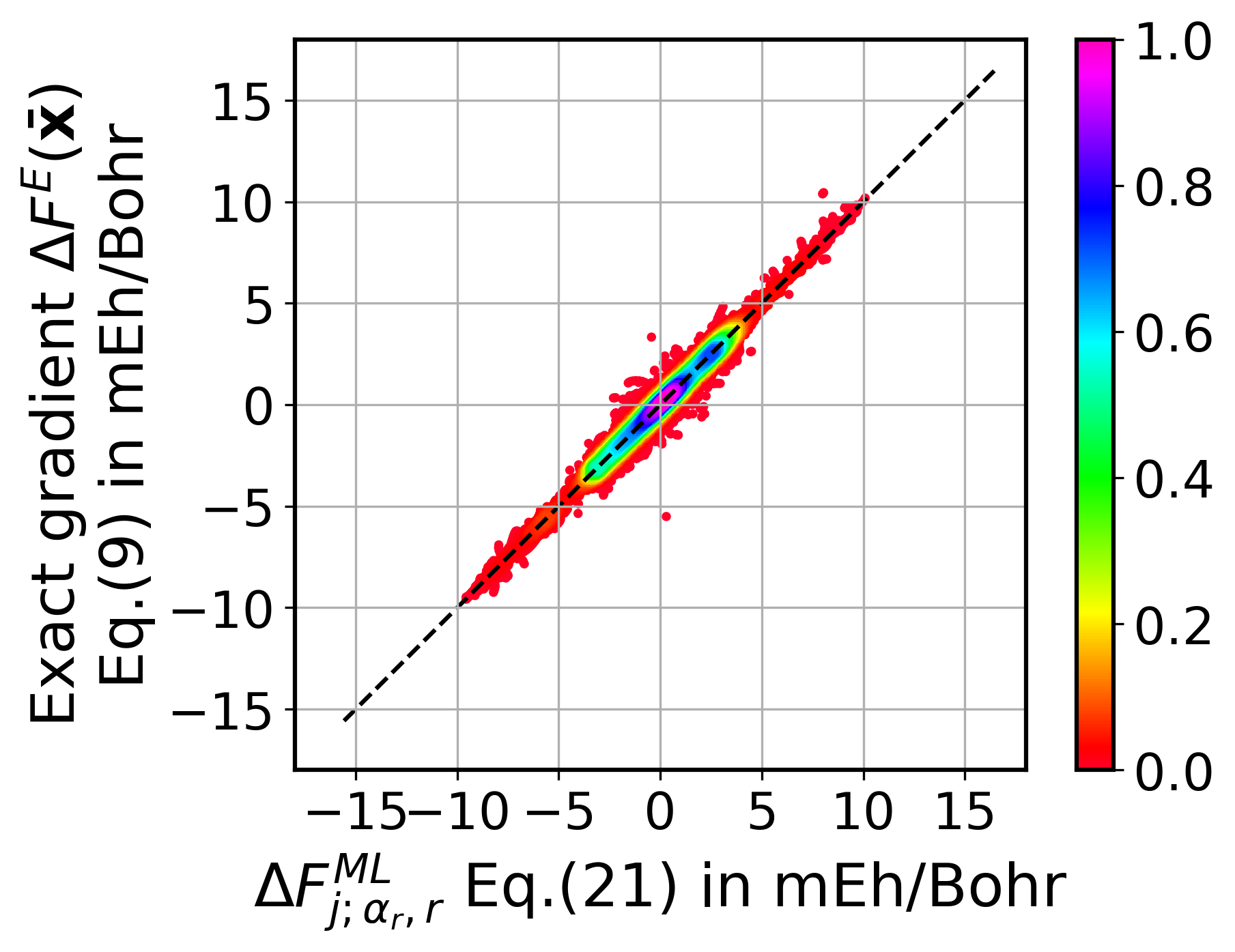}}
\subfigure[H$_7$O$_3^+$ with O-O-O angles between 90 and 145$^\circ$]{\includegraphics[width=0.23\textwidth]{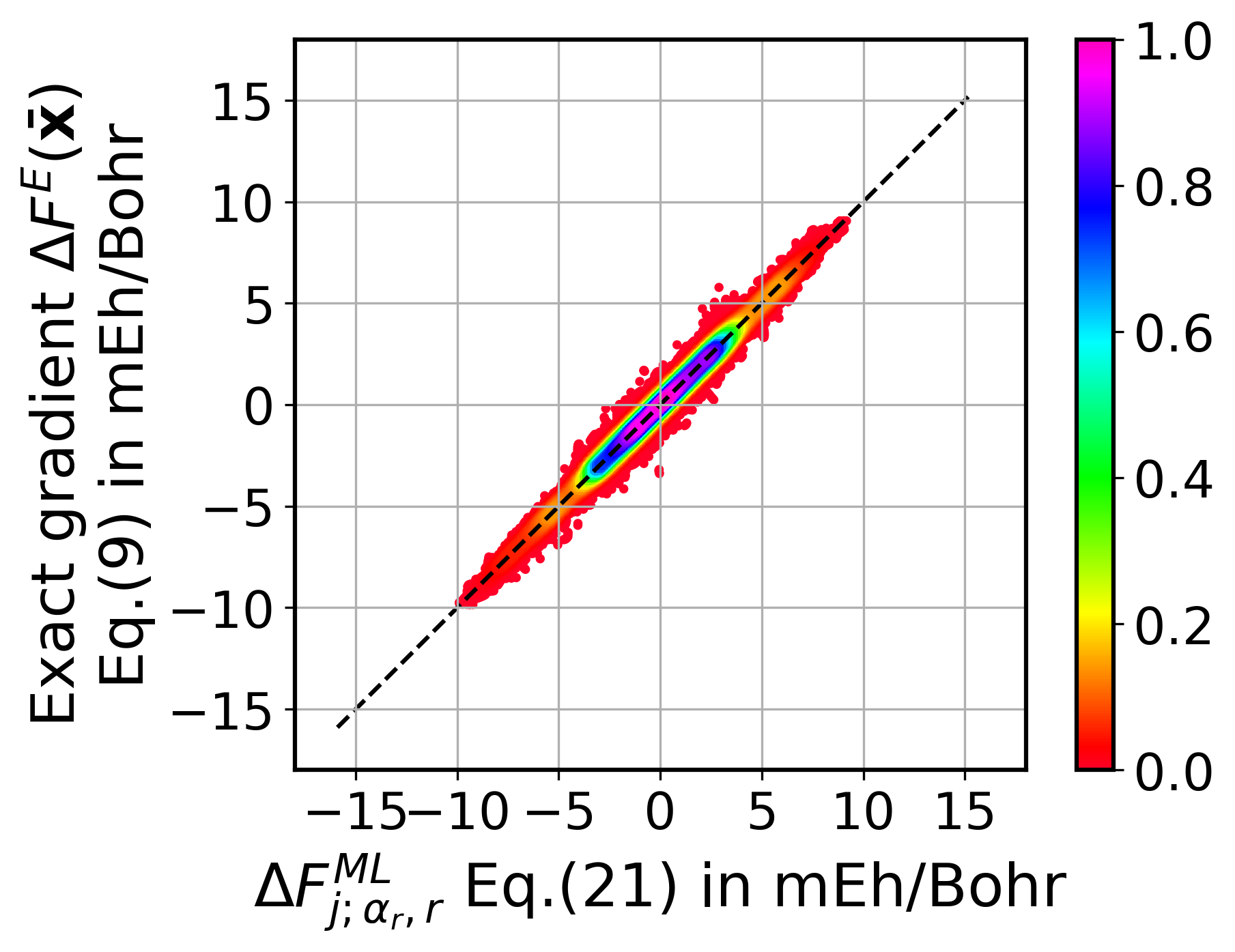}}
\subfigure[H$_7$O$_3^+$ with O-O-O angles $\geq$ 145$^\circ$]{\includegraphics[width=0.23\textwidth]{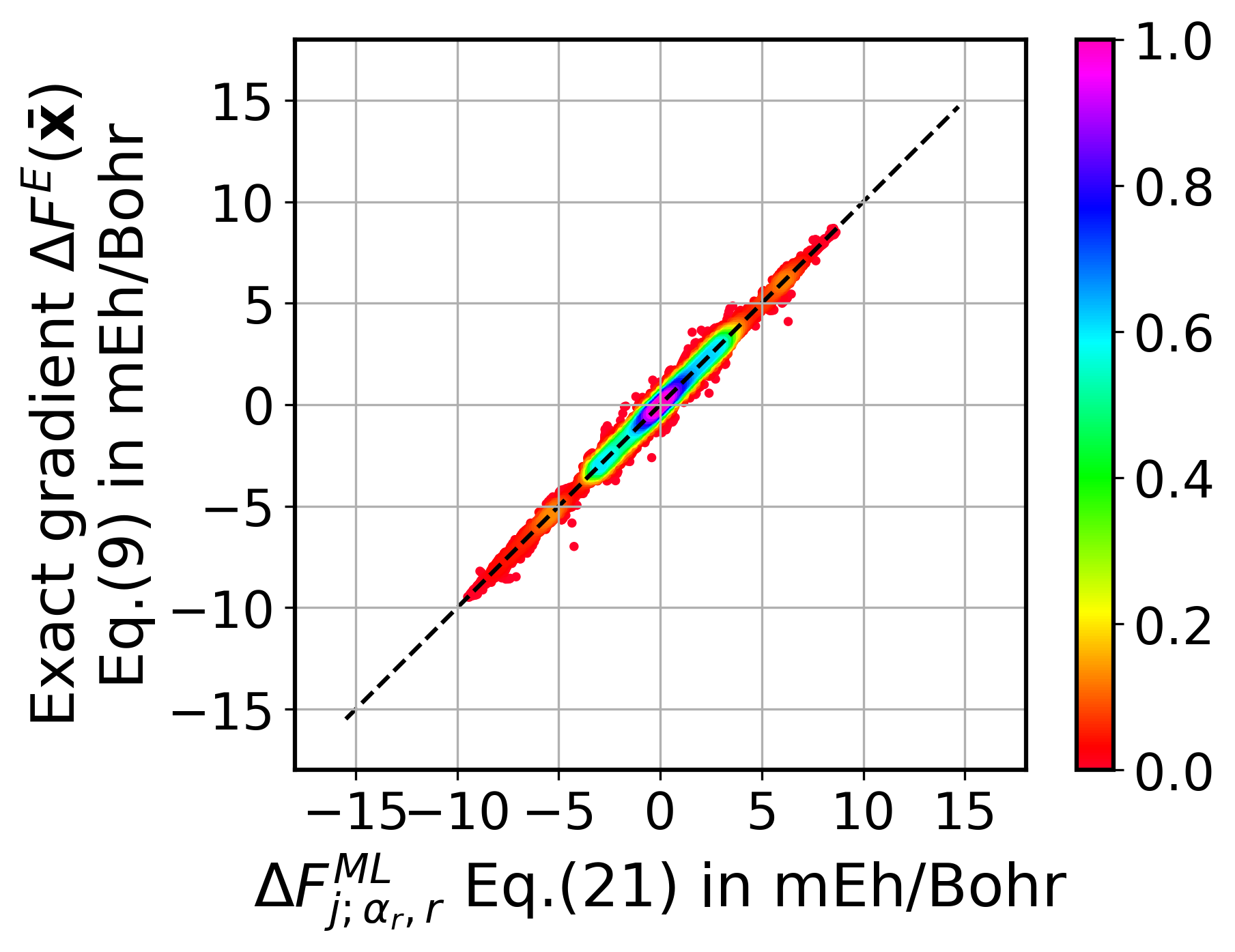}}

\caption{\label{fig_h7o3err} Distribution of fragment force errors for the given the range of oxygen-oxygen distance for $H_7O_3^+$.}
\end{figure}

In this section, we provide more detailed discussion about the geometrical variance on fragments and their corresponding model accuracy as discussed in section \ref{sec_frag}. Specifically, each type of fragments are divided into different groups based on the oxygen oxygen distance and the oxygen-oxygen-oxygen angles as shown in Figure \ref{fig:RDF} in Section \ref{sec_result}. For rank 1 simplexes $H_4O_2$ and $H_5O_2^+$, we divide them into three categories in Figures \ref{fig_h4o2err} and \ref{fig_h5o2err} based on the oxygen oxygen distance. For rank 2 simplexes $H_6O_3$ and $H_7O_3^+$, we divide them into three categories in Figures \ref{fig_h6o3err} and \ref{fig_h7o3err} based on the oxygen oxygen oxygen angles. 

As we can see in Figures \ref{fig_h4o2err} and \ref{fig_h5o2err}, even with only 10\% training, the model achieve high accuracy for a wide range of geometries with oxygen oxygen distance ranging from 2{\AA} to 7{\AA} as shown in the oxygen oxygen distribution plot in Figure \ref{fig:RDF}. The non hydrogen bounded dimers has smaller force component range but still can be accurately predicted with the same neural network model. This guarantees the transferability of models across a variety of different geometries and potentially applicable to different molecular systems.





\bibliographystyle{achemso-ssigrp}
\providecommand{\latin}[1]{#1}
\providecommand*\mcitethebibliography{\thebibliography}
\csname @ifundefined\endcsname{endmcitethebibliography}  {\let\endmcitethebibliography\endthebibliography}{}

\end{document}